\documentclass[12pt,letterpaper]{article}        

\usepackage{jheppub}
\pdfoutput=1
\usepackage[utf8]{inputenc}
\usepackage{amsmath,amssymb}
\usepackage{amsfonts}
\usepackage{slashed}
\usepackage{xcolor}

\usepackage{changes}
\usepackage{graphicx}

\begin{document}



\title{Diffusion and universal relaxation of holographic phonons}
\author[1,2]{Andrea Amoretti,}
\author[3]{Daniel Are\'an,}
\author[4]{Blaise Gout\'{e}raux,} 
\author[5]{and Daniele Musso}

\emailAdd{andrea.amoretti@ge.infn.it}
\emailAdd{daniel.arean@uam.es}
\emailAdd{blaise.gouteraux@polytechnique.edu}
\emailAdd{daniele.musso@usc.es}

\preprint{CPHT-RR018.042019 IFT-UAM/CSIC-19-55}

\affiliation[1]{Dipartimento di Fisica, Universit\`a di Genova,
via Dodecaneso 33, I-16146, Genova, Italy and I.N.F.N. - Sezione di Genova
}
\affiliation[2]{Physique  Th\'{e}orique  et  Math\'{e}matique  and  International  Solvay  Institutes  Universit\'{e}  Libre de Bruxelles, C.P. 231, 1050 Brussels, Belgium}
\affiliation[3]{Instituto de F\'\i sica Te\'orica UAM/CSIC,
	Calle Nicol\'as Cabrera 13-15, Cantoblanco, 28049 Madrid, Spain}
\affiliation[4]{CPHT, CNRS, Ecole polytechnique, IP Paris, F-91128 Palaiseau, France}
\affiliation[5]{Departamento de Física de Partículas, Universidade de Santiago de Compostela and
Instituto Galego de Física de Altas Enerxías (IGFAE), 15782 Santiago de Compostela, Galicia, Spain.}

\abstract{
In phases where translations are spontaneously broken, new gapless degrees of freedom appear in the low energy spectrum (the phonons). At long wavelengths, they couple to small fluctuations of the conserved densities of the system. This mixing is captured by new diffusive transport coefficients, as well as qualitatively different collective modes, such as shear sound modes. We use Gauge/Gravity duality to model such phases and analytically compute the corresponding diffusivities in terms of data {of the dual background black hole solution}. In holographic quantum critical low temperature phases, we show that these diffusivities are governed by universal relaxation of the phonons into the heat current when the dynamical critical exponent  $z>2$. Finally, we compute the spectrum of transverse collective modes and show that their dispersion relation matches the dispersion relation of the shear sound modes of the hydrodynamic theory of crystalline solids.
}

\maketitle

\tableofcontents
\section{Introduction}

One of the most celebrated results originating from Gauge/Gravity duality \cite{Maldacena:1997re,Ammon:2015wua,Zaanen:2015oix,Hartnoll:2016apf} is the computation of the shear viscosity of the $\mathcal N=4$ super Yang-Mills plasma at infinite $N$ {($N$ is the rank of the gauge group)} and 't Hooft coupling in terms of the entropy density of a dual black hole in anti de Sitter spacetime, $\eta=s/4\pi$ (in natural units) \cite{Policastro:2001yc}. This was conjectured to place a lower bound on the ratio $\eta/s\gtrsim 1/4\pi$ for strongly-coupled phases of matter \cite{Kovtun:2004de}. 
Recalling that the entropy density is proportional to the area of the bulk black hole, 
this relation is universal in the sense that the complicated dependence on the boundary sources is encapsulated as a 
simple combination of data defined at the black hole horizon \cite{Iqbal:2008by}.

In a relativistic plasma, which is described at long wavelengths by relativistic hydrodynamics (see \cite{Kovtun:2012rj} for a review of relativistic hydrodynamics), 
the shear viscosity controls the diffusion of transverse momentum $D_\perp=\eta/(\epsilon+p)$, 
where $\epsilon$ and $p$ are the energy density and pressure.

At nonzero density, relativistic hydrodynamics predicts the existence of another, incoherent diffusion constant $\sigma_0$,
associated to processes without any momentum drag. This diffusivity can also be computed in terms of horizon data \cite{Davison:2015taa,Davison:2018ofp,Davison:2018nxm}, 
as well as the thermal diffusivity that governs energy diffusion \cite{Blake:2017qgd}.

Taken together, these results point at an interesting link between universal low energy transport properties (diffusivities) and the dynamics close to black hole horizons.

These results have been extended to less symmetric cases describing inhomogeneous states, such as the so-called holographic lattices that break translations explicitly with a periodic potential \cite{Donos:2014yya,Donos:2015gia,Banks:2015wha}. At long distances, charge and energy diffuse in these inhomogeneous systems \cite{Hartnoll:2014lpa,Donos:2017gej}. The diffusivities are proportional to the dc thermoelectric conductivities of the system via Einstein relations, which relates them to horizon data under very general assumptions \cite{Donos:2017ihe}. 

More recent developments have considered the effect of breaking translations spontaneously in one or more spatial directions, which can be implemented either in a homogeneous\footnote{The broken translation generator then combines with an internal symmetry such that a diagonal combination is preserved.} \cite{Amoretti:2016bxs,Alberte:2017oqx,Amoretti:2017frz,Amoretti:2017axe} or an inhomogeneous way \cite{Ooguri:2010kt,Donos:2011bh,Donos:2013wia,Withers:2013loa,Jokela:2014dba,Jokela:2016xuy}. In either case, the incoherent conductivity and associated diffusivity of the boundary theory can be expressed in terms of horizon data {for thermodynamically stable phases} \cite{Amoretti:2017frz,Amoretti:2017axe,Donos:2018kkm,Gouteraux:2018wfe}. {For thermodynamically unstable phases, the incoherent conductivity also features an integral over the whole spacetime of some combination of background fields.}

The effective theory \cite{chaikin_lubensky_1995,Delacretaz:2017zxd} capturing the low energy dynamics of such states with spontaneous translation symmetry breaking contains more than the two characteristic diffusivities just mentioned, due to the presence of additional gapless degrees of freedom (the Goldstones of spontaneous translation symmetry breaking or in other words, the phonons) and their mixing with charge and energy fluctuations. The theory of Wigner crystal (WC) hydrodynamics is succinctly reviewed in section \ref{section:WChydro} and the diffusivities we are interested in are presented there. Then, in section \ref{section:holo}, we give some details of our holographic setup. {A first objective of this work is to extend previous analyses and show that all of these diffusivities can be expressed in terms of the background black hole solution, by a combination of data on the horizon and on the rest of the spacetime}. This is done in section \ref{section:unrelaxedDC}, and our main new results are equations \eqref{diffDCgamma1} and \eqref{diffDCxi}.

Positivity of entropy production in WC hydrodynamics places a bound on a combination of the diffusivities, see equation \eqref{EntropyConstraints}. The holographic diffusivities obey this bound, as we discuss in section \ref{section:bound}. Interestingly, this bound can be saturated at low temperatures. This obtains when the phonons relax into the heat current and leads to simple relations between the diffusivities. We expect this universal relaxation channel to be at play in generic states of matter at finite temperature. We give a criterion for the saturation of this bound as a function of the values of the scaling exponents characterizing the infra-red fixed point of the system and the irrelevant deformations away from it. Universal relaxation of phonons into a hydrodynamic operator has also been recently discussed in the context of the melting of the field-induced Wigner solid in \cite{Delacretaz:2019}.

In section \ref{sec:transfluc}, we conclude with an analysis of the spectrum of the system at nonzero wavevector by computing the longest-lived pair of  transverse quasinormal modes (QNMs) of the dual black hole. We show that they obey the expected dispersion relation from WC hydrodynamics. At low wavevector and intermediate temperatures, there is a pair of shear sound modes, characteristic of the long wavelength dynamics of solids. However, either at very low or very high temperatures, these modes collide on the imaginary frequency axis and become pseudo-diffusive, with a purely imaginary gap controlled by a ratio of thermodynamic data and diffusivities. Both collisions occur in the hydrodynamic regime and the dispersion relation of the modes is well-captured by the hydrodynamic dispersion relation, see equation \eqref{hydroapp}.  These results were originally presented in \cite{Amoretti:2018tzw}, but have been moved and expanded upon in the present work for clarity.

We give a number of technical digressions as well as details of our numeric scheme in some appendices.

\section{Review of Wigner crystal hydrodynamics \label{section:WChydro}}

We first recap the main features of {two-dimensional} isotropic Wigner crystal hydrodynamics. More details can be found in \cite{chaikin_lubensky_1995,Delacretaz:2017zxd}.
As translations are broken spontaneously along both spatial directions, the usual conserved densities (energy, charge, momentum) need to be coupled to two Goldstone modes, $\varphi_i$, $i=x,y$. The free energy is supplemented by terms capturing the effect of the Goldstones:
\begin{equation}
f=\frac12 K|q\cdot\varphi_q|^2+\frac12Gq^2|\varphi_q|^2\,.
\end{equation}
$K$ and $G$ are the bulk and shear moduli, and characterize the stiffness of phase fluctuations around the ordered state. It is convenient to parameterize the Goldstones by their longitudinal and transverse contributions, $\lambda_{\parallel}=\nabla\cdot\varphi$ and $\lambda_\perp=\nabla\times\varphi$. The corresponding sources $s_{\parallel,\perp}$ are defined by requiring that $\lambda_{\parallel,\perp}=\delta f/\delta s_{\parallel,\perp}$.

To leading order in gradients and keeping only linear terms, the Goldstones obey the following `Josephson' relations
\begin{equation}
\label{Josephson}
\begin{split}
\partial_t \lambda_\parallel=&\nabla\cdot v+\gamma_1 \nabla^2\mu+\gamma_2 \nabla^2 T+\frac{\xi_\parallel}{K+G}\nabla^2 s_\parallel+\dots\,,\\
\partial_t \lambda_\perp=&\nabla\times v+\frac{\xi_\perp}{G}\nabla^2 s_\perp+\dots\,,
\end{split}
\end{equation}
where $v$ is the velocity, $\mu$ the chemical potential, and $\gamma_{1,2}$ and $\xi_{\parallel,\perp}$ are diffusive transport coefficients.

These equations are supplemented by current, heat and momentum conservation equations (energy can be traded for entropy to linear order):
\begin{equation}
\label{Conseqhydro}
\partial_t \rho+\nabla\cdot j=0\,,\quad \partial_t s+\nabla\cdot (j_q/T)=0\,,\quad\partial\pi^i+\nabla_j T^{ji}=0\,,
\end{equation}
together with the constitutive relations
\begin{equation}
\begin{split}
j=&\rho v-\sigma_o\nabla\mu-\alpha_o\nabla T-\gamma_1\nabla s_\parallel+\dots\,,\\
j_q/T=&s v-\alpha_o\nabla\mu-(\bar\kappa_o/T)\nabla T-\gamma_2\nabla s_\parallel+\dots\,,\\
T^{ij}=&\delta^{ij}\left(p+(K+G)\nabla\cdot\varphi\right)+2G\left[\nabla^{(i}\varphi^{j)}-\delta^{ij}\nabla\cdot\varphi\right]-\eta\left(2\nabla^{(i} v^{j)}-\delta^{ij}\nabla\cdot v\right)+\dots
\end{split}
\end{equation}
The underlying conformal symmetry of the holographic setup implies that the stress-energy tensor is traceless, which sets the bulk viscosity to zero.

The hydrodynamic retarded Green's functions at nonzero frequency and wavevector are derived by following the Kadanoff-Martin procedure \cite{1963AnPhy..24..419K,Kovtun:2012rj}:
\begin{equation}
\label{GABKM}
G^R_{AB}(\omega,q)=M_{AC}\left[\left(i\omega-M\right)^{-1}\right]_{CD}\chi_{DB}
\end{equation}
where the vevs $A,B=(\delta\rho,\delta s,\pi_{\parallel},\lambda_\parallel,\pi_\perp,\lambda_\perp)$ and the corresponding sources are\\ $(\delta\mu,\delta T,v_{\parallel}, s_\parallel,v_\perp, s_\perp)$. $M$ is the matrix
\begin{equation}\label{M}
M_{AB}=\left(\begin{array}{cccccc}
\sigma_o q^2&\alpha_o q^2&i \rho q&\gamma_1 q^2&0&0\\
\alpha_o q^2&\frac{\bar\kappa_o}{T}q^2 & i s q & \gamma_2 q^2 & 0 & 0 \\
i \rho q & is q&\eta q^2&iq&0&0\\
\gamma_1 q^2&\gamma_2 q^2&i q&\frac{\xi_\parallel}{K+G}q^2&0&0\\
0&0&0&0&\eta q^2&i q\\
0&0&0&0&iq&\frac{\xi_\perp}{G}q^2
\end{array}\right)\,.
\end{equation}
Relativistic symmetry of the holographic setup enforces that the momentum and energy current densities must be equal $\pi=j_e$, which places constraints on the transport coefficients:
\begin{equation}
\label{RelHydro}
\alpha_o=-\frac\mu{T}\sigma_o\,,\quad \bar\kappa_o=\frac{\mu^2}{T}\sigma_o\,,\quad \gamma_2=-\frac{\mu}T\gamma_1\,.
\end{equation}
Observe that this also means that the heat current $j_q\equiv j_e-\mu j=\pi-\mu j$.
Finally, the susceptibility matrix is\footnote{We have set to zero off-diagonal terms $\chi_{\rho\lambda_\parallel}$ and $\chi_{s\lambda_\parallel}$. As pointed out in appendix A of \cite{Delacretaz:2017zxd}, they affect the dispersion relation of modes in the longitudinal sector, but are not important here.}
\begin{equation}
\chi_{AB}=\left(\begin{array}{cccccc}
\chi_{\rho\rho}&\chi_{\rho s}&0&0&0&0\\
\chi_{\rho s}&\chi_{ss} & 0 &0 & 0 & 0 \\
0 & 0&\chi_{\pi\pi}&0&0&0\\
0&0&0&\frac{1}{K+G}&0&0\\
0&0&0&0&\chi_{\pi\pi}&0\\
0&0&0&0&0&\frac{1}{G}
\end{array}\right)\,.
\end{equation}

Positivity of entropy production can be ensured by requiring all the eigenvalues of the matrix $M$ to be positive \cite{Delacretaz:2017zxd},
which leads to the following constraints:
\begin{equation}
\label{EntropyConstraints}
\eta\geq0\,,\qquad \sigma_o\geq0\,,\qquad \gamma_1^2\leq{\frac{\sigma_o\xi_{\parallel}}{K+G}}\,.
\end{equation}
Using \eqref{GABKM} and identities between the Green's functions stemming from \eqref{Conseqhydro}, we obtain in the $q\to0$ limit
\begin{eqnarray}
\label{KuboUnrelaxed}
&G^R_{T^{xy}T^{xy}} =G-i\omega\eta\,,\\
\label{KuboUnrelaxedlambda}
&G^R_{j j} = \frac{\rho^2}{\chi_{\pi\pi}}-i\omega\sigma_o\,,\quad \quad
G^R_{j\pi_\parallel}=\rho\,, \quad \quad
G^R_{j\varphi_\parallel}=-\gamma_1-\frac{i \rho}{\chi_{\pi\pi}\,\omega}\,,\\
\label{KuboUnrelaxedlambda2}
&G^R_{\pi_\parallel\varphi_\parallel}=G^R_{\varphi_\parallel\pi_\parallel}=\frac{i}{\omega}\,,\quad
G^R_{\varphi_\parallel\varphi_\parallel}=\frac{1}{\chi_{\pi\pi}\,\omega^2}
-\frac{\xi_\parallel}{K+G}\frac{i}{\omega}\,,\quad G^R_{\varphi_\perp\varphi_\perp}=
\frac{1}{\chi_{\pi\pi}\,\omega^2}-\frac{\xi_\perp}{G}\frac{i}{\omega}\,.
\end{eqnarray}
We defined {\it e.g.}
\begin{equation}
G^R_{\varphi_i \varphi_j}=\frac{q_i q_j}{q^4}G^R_{\lambda_\parallel\lambda_\parallel}+\left[\delta_{ij}-\frac{q_i q_j}{q^2}\right]\frac{G^R_{\lambda_\perp\lambda_\perp}}{q^2}\,,
\end{equation}
and 
\begin{equation}
{q^2G^R_{\varphi_\parallel\varphi_\parallel}=G^R_{\lambda_\parallel\lambda_\parallel}\,,\quad  q^2G^R_{\varphi_\perp\varphi_\parallel}=G^R_{\lambda_\perp\lambda_\perp}}\,.
\end{equation}
In this $q=0$ limit, for an isotropic crystal, $G^R_{\lambda_\parallel\lambda_\parallel}=G^R_{\lambda_\perp\lambda_\perp}$, 
since there should be no distinction between the longitudinal and tranverse phonons. This leads to the constraint
\begin{equation}
\label{Xdef}
\frac{\xi_\parallel}{K+G}=\frac{\xi_\perp}{G}\equiv X\,.
\end{equation}

\section{Holographic model \label{section:holo}}

\subsection{Setup}

We consider the holographic model
\begin{equation}\label{action}
S=\int d^{4}x\,\sqrt{-g}\left[R-\frac12\partial\phi^2-V(\phi)-\frac14 Z(\phi)F^2-\frac12\sum_{I=1}^{2}Y(\phi)\partial \psi_I^2\right],
\end{equation}
with the scalar couplings behaving near the AdS boundary, {i.e. in the small $\phi$ limit, as}
\begin{equation}
 \label{eq:UvExpansionCouplings}
V_{uv}(\phi)=-6-\phi^2+O(\phi^3)\,,\quad  Z_{uv}(\phi)=1+O(\phi)\,,\quad Y_{uv}(\phi)=\phi^2+O(\phi^3)\,.
\end{equation}

The model \eqref{action} enjoys a global shift symmetry $\psi_I\mapsto\psi_I+c_I$.
 In this work, we will be interested in states that break translations homogeneously \cite{Donos:2013eha,Andrade:2013gsa},\footnote{See \cite{Musso:2018wbv} for a recent field theoretic investigation of such states.} with 
 \begin{equation}
 \label{eq:backpsi}
 \psi_I=k x^I\,,\quad x^I=\{x,y\}\,.
 \end{equation}
This Ansatz breaks the shift symmetry as well as spacetime translations to a diagonal U(1).
As a consequence, the background metric, scalar $\phi$ and gauge field only depend on the holographic radial coordinate, 
\begin{equation}
  ds^2=-D(r)dt^2+B(r) dr^2+C(r)(dx^2+dy^2),\quad \phi=\phi(r),\quad A=A(r)dt\,.
\end{equation}
$t,x,y$ are the time and space coordinates of the dual field theory. Given \eqref{eq:UvExpansionCouplings} and \eqref{eq:backpsi}, the dual boundary theory is deformed by two complex scalar operators $\Phi_I\simeq \phi e^{i\psi_I}$\,. These deformations break translations through the spatial dependence of the phases $\psi_I$.

As explained in \cite{Amoretti:2016bxs,Amoretti:2017frz,Amoretti:2017axe}, 
whether translations are broken explicitly or spontaneously in the boundary theory depends on the asymptotic behaviour of the scalar $\phi$ near the AdS boundary. {For our choice of scalar potential, $\phi$ decays towards the boundary as}
\begin{equation}
\phi(r\to0)=\lambda r+\phi_{(v)} r^{2}+O(r^3)\,.
\end{equation}
In the usual quantization scheme, $\lambda$ is the source of the operator dual to the bulk field $\phi$ and $\phi_{(v)}$ is related to its vev. 
If $\lambda=0$, the breaking is spontaneous. If $\lambda\neq0$, it is explicit. In this work we will only consider the spontaneous case.

The condensation of the order parameter itself
({\it i.e.} the phase transition between the normal and ordered phase) is not captured by the model. This is typically mediated by an instability towards a spatially modulated, 
inhomogeneous phase which minimizes the free energy, 
\cite{Donos:2013wia,Withers:2013loa,Withers:2014sja,Jokela:2016xuy}.\footnote{This can also be realized in models breaking 
translations homogeneously \cite{Andrade:2017cnc,Amoretti:2017frz,Amoretti:2017axe}.} 
Instead, the holographic model \eqref{action} with zero scalar source $\lambda=0$ directly describes the low energy dynamics of the phonons in the  ordered phase, 
as we shall demonstrate in the remainder of this work.\footnote{Minimizing the free energy would lead to $k=0$, 
{\it i.e.} no breaking of translations at all, \cite{Amoretti:2017frz,Amoretti:2017axe,Donos:2018kkm}. This can be remedied by turning on higher-derivative terms in the action with small couplings  \cite{Amoretti:2017frz,Amoretti:2017axe}.}
As we show below, the fluctuations of the bulk fields $\psi_I$ are dual to the phonons. 
The bulk global symmetry should then be understood as encoding the shift symmetry of the NG bosons, not as a symmetry of the fundamental UV theory.\footnote{A very similar model was considered in \cite{Iqbal:2010eh} to study holographically spin density waves.}

Finite temperature, finite density states are modeled by charged black holes in the bulk, which implies the existence 
of a regular black hole horizon at $r=r_h$. In  the rest of this work, we use a subscript $h$ to denote quantities evaluated at $r=r_h$. 
The temperature $T$ and the entropy density $s$ are given by:
\begin{equation}
s=4 \pi C(r_h) \ , \qquad T=\frac{1}{4 \pi }\left.\sqrt{-\frac{B'(r)D'(r)}{B(r)^2}}
\right|_{r=r_h}\ ,
\end{equation}
with the following near-horizon expansion
\begin{equation}
\label{NearHorizon}
\begin{split}
&ds^2=-4\pi T(r_h-r)dt^2+\frac{dr^2}{4\pi T(r_h-r)}+\frac{s}{4\pi}(dx^2+dy^2)+\dots \ ,\\
&A_t=A_h(r_h-r)+\dots\ , \qquad \phi=\phi_h+\dots\ .
\end{split}
\end{equation}

We will also be interested in the low temperature behaviour $T\ll\mu$ (where $\mu$ is the chemical potential) of these translation-breaking black holes. 
More precisely, we would like to study the interplay between spontaneous translation symmetry breaking and quantum criticality. 
To do so, we will assume that the scalar $\phi$ has a runaway behaviour as $T\to0$ {and $r\to+\infty$ (or equivalently that the horizon value of the scalar diverges at low temperature)}, and that the scalar couplings behave for large $\phi$ as
\begin{equation}
\label{IRscalarcouplings}
V(\phi\to\infty)=V_0 e^{-\delta\phi}\ ,\quad Z(\phi\to\infty)=Z_0 e^{\gamma\phi}\ ,\quad Y_{IR}=Y(\phi\to\infty) e^{\nu\phi}\,.
\end{equation}
The near-extremal, near-horizon geometry of the black hole is then described by a Lifshitz, hyperscaling-violating solution 
\begin{equation}
\label{skaska}
\begin{split}
& ds^2 = \xi^\theta \left[ -f(\xi)\frac{dt^2}{\xi^{2z}} + \frac{L^2 d\xi^2}{\xi^2f(\xi) } + \frac{d\vec{x}^2}{\xi^2}\right], \quad f(\xi)=1-\left(\frac{\xi}{\xi_h}\right)^{2+z-\theta}\,,\\
& 
 A = A_o\, \xi^{\theta-2 - z} dt\ ,\quad
\psi_I = k\delta_{Ii} x^i\,,\quad \phi=\kappa\log \xi\,.
 \end{split}
\end{equation}
{$\xi$ is a radial coordinate valid close to $r_h$, simultaneously taking the small temperature limit $T\ll\mu$.}
$\xi_h$ is the location of a Killing event horizon, with the associated Hawking temperature $T\sim \xi_h^{-1/z}$ and Hawking-Bekenstein entropy $s\sim \xi_h^{\theta-2}$. Combining both formul\ae\ the entropy density scales as $s \sim T^{\frac{2-\theta}{z}}$. {More details of these solutions can be found in \cite{Gouteraux:2014hca} or more recently \cite{Davison:2018nxm}. Here, we simply recall that in general the couplings appearing in \eqref{IRscalarcouplings} are related to the scaling exponents characterizing the solution as
\begin{equation}
\label{irexponents}
\kappa\delta=\theta\,,\qquad \kappa\gamma=4-\theta-2\Delta_{A_o}\,,\qquad \kappa\nu=2\Delta_k-2\,.
\end{equation}
The exponents $\Delta_{A_o,k}\leq0$ are the scaling dimension of sources of irrelevant operators breaking particle/hole symmetry or translations of the IR $T=0$ critical solution.
}

For concreteness, in our numerical calculations we work with 
\begin{equation}
\label{eq:potentials}
V(\phi)=-6\cosh(\phi/\sqrt3)\,, \quad Z(\phi)=\exp(-\phi/\sqrt3)\,,\quad Y(\phi)=(1-\exp\phi)^2\,.
\end{equation}
This implies that the critical solution \eqref{skaska} is characterized by $\theta\to-\infty$, $z\to+\infty$, $-\theta/z=1$ and $\Delta_{A_o,k}=0$. As a consequence, it has a vanishing entropy density $s\sim T$. Our results can easily be generalized to other scalings. The numerical results presented in this paper are for $k/\mu=0.1$ and $\lambda=0$.

\subsection{Holographic renormalization}

Holographic renormalization for the spontaneous case was explained in \cite{Amoretti:2017frz}, to which we refer for details. 
The salient features are the following. The UV expansion of the fields $\psi_I$ are modified compared to that of ordinary massless scalars 
due to their coupling to the scalar $\phi$. If $\phi$ is not sourced $\lambda=0$, then close to the AdS boundary $r\to0$
\begin{equation}
\psi_I=\frac{\psi_{I,(-1)}}{r}+\psi_{I,(0)}+O(r)=\frac1{r^2}\left(\psi_{I,(-1)}r+\psi_{I,(0)}r^2+O(r^3)\right)\,,
\end{equation}
where we have used that {the operator dual to} $\phi$ has dimension $\Delta=2$ due to \eqref{eq:potentials}. This makes clear that {the operators dual to} the $\psi_I$ have the same scaling dimension as {the operator dual to} $\phi$ and that the whole scalar sector {of deformations of the UV CFT} is better understood 
as complex scalar operators {dual to bulk complex scalar fields} $\Phi_I\simeq\phi e^{i\psi_I}$ {for small $\phi$ (ie close to the AdS boundary)}. {$\phi$ and the $\psi_I$ are respectively the modulus and phase of the complex scalar, and inherit its scaling dimension}. {See \cite{Amoretti:2017frz} for details}. Going back to our background Ansatz \eqref{eq:backpsi}, we see that indeed 
this corresponds to spontaneous translation breaking, since the source term $\psi_{(-1)}$ is not turned on.
These powers can be generalized easily to the case $\Delta\neq2$ and spatial dimension $d\neq2$.

Near the boundary, the metric and gauge field behave in Fefferman-Graham gauge as
\begin{equation}
\begin{split}
&D(r)=\frac1{r^2}\left(1+d_{(3)} r^3+O(r^4)\right)\ , \quad
C(r)=\frac1{r^2}\left(1-\frac{d_{(3)}}2 r^3+O(r^4)\right), \\
&B(r)=\frac1{r^2}\ , \hspace{4.2cm} A(r)=\mu - \rho r+O(r^3) \,,
\end{split}
\end{equation}
where subleading coefficients are fixed in terms of, $\rho$, $\phi_{(v)}$ and $d_{(3)}$.
This allows to read off the one-point correlation functions of the dual field theory, after computing the on-shell action renormalized by suitable counterterms \cite{deHaro:2000vlm}.  We find \cite{Amoretti:2017frz}
 \begin{equation}
\label{stressenergytensor}
\begin{split}
\langle T^{tt} \rangle= \epsilon=- 3 d_{(3)} \,, \quad &
\langle T^{xx} \rangle=\langle T^{yy} \rangle=- \frac32 d_{(3)}
 \\
&\langle J^{t} \rangle=\rho \ ,\quad
\langle O_\phi \rangle=\phi_{(v)}\,.
\end{split}
\end{equation}
Another useful relation is
\begin{equation}\label{smarr}
s T=- \mu \rho -\frac{9 d_{(3)}}{2}-k^2\int_0^{r_h} dr\, \sqrt{BD}Y\,,
\end{equation}
which results from evaluating the radially conserved bulk expression 
\begin{equation}\label{radial1}
\left[\rho A(r)+\frac{C^2(r)}{\sqrt{B(r)D(r)}}\left(\frac{D(r)}{C(r)}\right)'
-k^2\int_r^{r_h}d\tilde{r}\, \sqrt{BD}Y\right]'=0\,.
\end{equation}
both at the horizon and at the boundary. This is the Noether charge associated to the bulk time-like Killing vector \cite{Papadimitriou:2005ii,Gouteraux:2018wfe}. {By evaluating the on-shell action for the background solution, we also find that $p=sT+\mu\rho-\epsilon$.}

From the analysis above,  
$\chi_{\pi\pi}$ was obtained exactly in $k$ \cite{Amoretti:2017frz,Amoretti:2017axe}:
\begin{equation}
\label{chiPP}
\chi_{\pi\pi}=sT+\mu\rho+k^2 I_Y\,,\quad I_Y=\int_0^{r_h} dr\,\sqrt{BD}Y\,.
\end{equation}

In this paper, we will only be interested in the transverse sector of fluctuations $\delta g_{t}^x(r,t,x)= \delta h_t^x(r) e^{-i\omega t+i q y}$, $\delta g_{x}^y(r,t,x)=\delta h_x^y(r) e^{-i\omega t+i q y}$, $\delta g_{r}^x(r,t,x)=\delta h_r^x(r) e^{-i\omega t+i q y}$, $\delta a_x(r,t,x)=\delta a_x(r) e^{-i\omega t+i q y}$ and $\delta \psi_x(r,t,x)=\delta \psi_x(r) e^{-i\omega t+i q y}$. We have used the homogeneity of our Ansatz to expand the fluctuations in plane waves. In the radial gauge $\delta g_r^x=0$, and setting $q=0$ for now, their boundary expansions are
\begin{equation}
\label{flucUVexpsp}
\begin{split}
&\delta a_x(r)=\delta a_{(0)}+\delta a_{(1)}r+\dots\,,\\
&\delta h_t^x(r)=\delta h_{1,(0)}+\delta h_{1,(3)}r^3+\dots\,,\\
&\delta h_x^y(r)=\delta h_{2,(0)}+\delta h_{2,(3)}r^3+\dots\,,\\
&\delta \psi_x(r)=\frac{1}r\left(\delta \psi_{(-1)}+\delta \psi_{(0)}r+\dots\right)\,,
\end{split}
\end{equation}
where $(\delta a_{(0)},\delta h_{1,(0)},\delta h_{2,(0)},\delta \psi_{(-1)})$  are sources and $(\delta a_{(1)},\delta h_{1,(3)},\delta h_{2,(3)},\delta \psi_{(0)})$ are vevs.

Then, the renormalized action at quadratic order in the fluctuations reads: 
\begin{equation}\label{renspontaneous}
\begin{split}
S_{\rm ren}^{(2)}=\int d\omega &\left[\delta a_{(0)}\delta a_{(1)}-\rho\delta a_{(0)} \delta h_{1,(0)}-\frac32\delta h_{1,(0)}\delta h_{1,(3)}\right.\\
&\left.
+\frac32d_{(3)}(\delta h_{1,(0)})^2+(\phi_{(v)})^2\delta \psi_{(-1)}\delta \psi_{(0)}-\frac32\delta h_{2,(0)}\delta h_{2,(3)}+\frac32d_{(3)}(\delta h_{2,(0)})^2\right].
\end{split}
\end{equation}

\subsection{AC boundary correlators \label{section:acbdycorr}}

The Goldstone modes can be identified by acting on the background with the Lie derivative along $\partial/\partial_{\vec x}$. It leaves all fields invariant except the $\psi_I$'s. This confirms that phonon dynamics are captured by the fluctuations $\delta\psi_I$, and by computing the retarded Green's functions in the limit $\omega\neq0$, $q=0$, we will verify that they take the expressions predicted by WC hydrodynamics \eqref{KuboUnrelaxed}, \eqref{KuboUnrelaxedlambda} and \eqref{KuboUnrelaxedlambda2}.

At zero density, we can solve the equations analytically in the low frequency limit and confirm that we correctly obtain the zero density hydrodynamic correlators \eqref{KuboUnrelaxedlambda2} in appendix \ref{sec:bdycorrzerodensitysp}.
At nonzero density, we made this check numerically, and we describe our numerical methods in appendix \ref{app:numerics}.

The outcome of this analysis is that we identify the boundary phonon and its source along the direction $i={x,y}$ as 
\begin{equation}
\label{relPhononpsi}
\varphi_i=\frac{\delta\psi_{i,(0)}}{k(\phi_{(v)})^2}\,,\quad \delta s_{\varphi,i}=k(\phi_{(v)})^2\delta\psi_{i,(-1)}\,.
\end{equation}
Suppressing the spatial index, they appear in the on-shell action \eqref{renspontaneous} as
\begin{equation}\label{renspontaneousphi}
\begin{split}
S_{\rm ren}^{(2)}=\int d\omega &\left[\delta a_{(0)}\delta a_{(1)}-\rho\delta a_{(0)} \delta h_{1,(0)}-\frac32\delta h_{1,(0)}\delta h_{1,(3)}\right.\\
&\left.
+\frac32d_{(3)}(\delta h_{1,(0)})^2+(\phi_{(v)})^2 \varphi\delta s_\varphi-\frac32\delta h_{2,(0)}\delta h_{2,(3)}+\frac32d_{(3)}(\delta h_{2,(0)})^2\right].
\end{split}
\end{equation}

Once we have determined the on-shell action for the fluctuations
$\delta a_x$, $\delta \psi_x$, $\delta g_t^x$, it is easy to compute
the boundary correlators encoded in those bulk fluctuations~\cite{Kaminski:2009dh}.
This entails numerically solving the relevant equations of motion as we
describe in appendix \ref{subapp:flucs}.
We then verify that the zero frequency limit of those correlators
matches the hydrodynamic expressions \eqref{KuboUnrelaxedlambda} and \eqref{KuboUnrelaxedlambda2} 
for the retarded Green's functions $G^R_{AB}(\omega\to0,q=0)$. $A,B={j^x,\pi^x,\varphi_x}$, where $\chi_{\pi\pi}$ is 
given by \eqref{chiPP}. In section \ref{section:unrelaxedDC}, we also obtain expressions for the diffusivities $\sigma_o$, $X$ and $\gamma_1$
(introduced in \eqref{M}, \eqref{Xdef}) which give an excellent match to the correlators. Their expressions are given in \eqref{diffDCsigmao}-\eqref{diffDCxi}.

\subsection{Shear modulus and shear viscosity  \label{app:shear}}

We will now evaluate the shear modulus and the shear viscosity holographically using the low frequency shear correlator \eqref{KuboUnrelaxed}. 
The low frequency behavior of the shear correlator can be computed analytically in a standard way, following \cite{Lucas:2015vna,Davison:2015taa,Hartnoll:2016tri} (see \cite{Alberte:2017cch,Alberte:2017oqx,Baggioli:2018bfa} for computations of the shear correlator in spontaneous holographic setups, \cite{Alberte:2015isw,Alberte:2016xja,Burikham:2016roo,Ciobanu:2017fef} in explicit setups). The equation of motion for the metric perturbation $h_{x}^y= h_2(r) e^{-i\omega t}$:
\begin{equation}
\label{hxyeom}
 \left(\sqrt{\frac{D}{B}} C h_{2}' \right)' - k^2 Y(\phi) \sqrt{BD} h_{2} + \omega^2 \sqrt{\frac{B}{D}} C h_{2}=0\ .
\end{equation}
Setting $\omega=0$, there are two linearly independent solutions, one regular $h^{(reg)}_{2}$ at the horizon and one singular $h^{(sing)}_{2}$. The singular solution can be expressed from the regular one
\begin{equation}
h^{(sing)}_{2}=h^{(reg)}_2\int^r_{0}\frac{\sqrt{B}}{\sqrt{D}C \left(h^{(reg)}_{2}\right)^2}\ ,
\end{equation}
and behaves near the horizon as
\begin{equation}
h^{(sing)}_{2}(r_h)\sim-\frac{\log(r_h-r)}{s T h_2^{(reg)}(r_h)}+\textrm{finite}.
\end{equation}
When $k\neq0$, the regular solution can only be found perturbatively in $k$:
\begin{equation}
h^{(reg)}_{2}(r)=1-k^2 \int_{0}^{r}dr_1\sqrt{\frac{B}D}\frac{1}{C}\int^{r_h}_{r_1}  \sqrt{BD}Y+O(k^4)\,.
\end{equation}
At the horizon we should impose ingoing boundary conditions
\begin{equation}
h_2(r)=h^{(reg)}_{2}(r_h)e^{-\frac{i\omega}{4\pi T}\log(r_h-r)}+\dots \ ,
\end{equation}
which at small frequencies reads
\begin{equation}
\label{hxyingoinglowomega}
h_2(r)=h^{(reg)}_{2}(r_h)\left(1-\frac{i\omega}{4\pi T}\log(r_h-r)\right)+O(\omega^2) \ .
\end{equation}

\begin{figure}
\begin{tabular}{cc}
\includegraphics[width=0.45\textwidth]{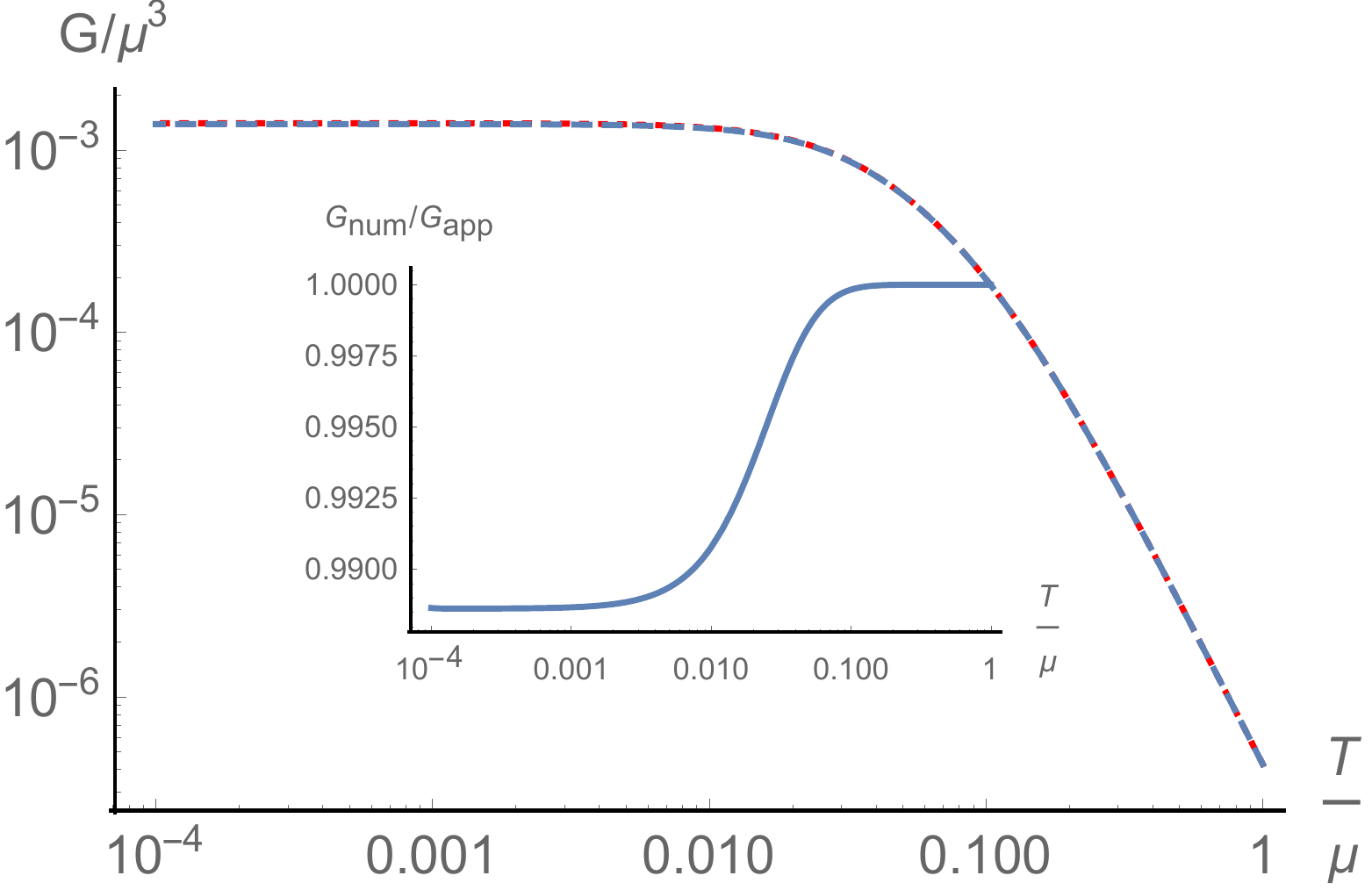}&\includegraphics[width=0.45\textwidth]{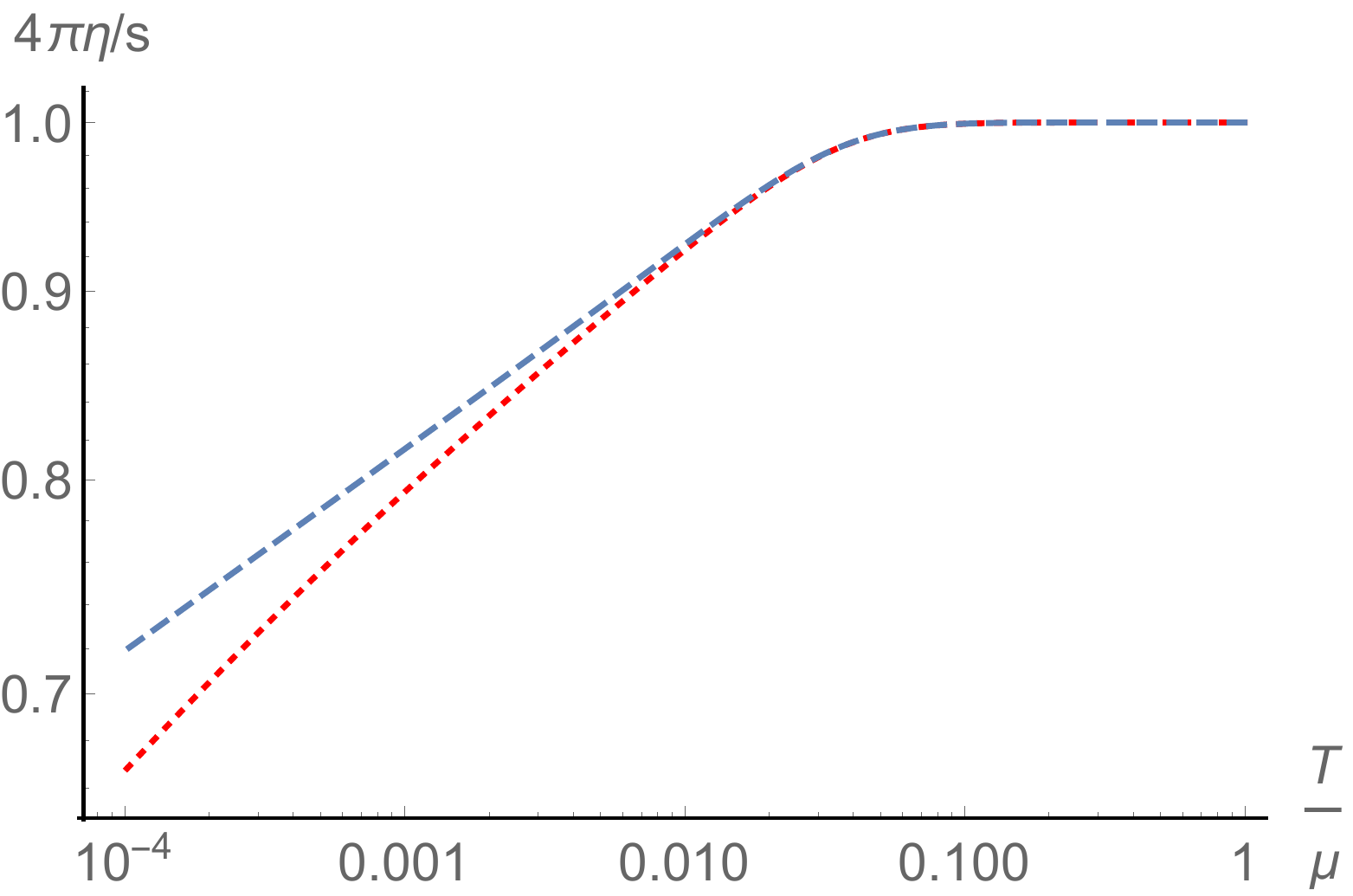}
\end{tabular}
\caption{In both plots, the blue dashed curve is the exact numerical result from the Kubo formul\ae\ \eqref{KuboUnrelaxed}; the red dotted curve is the approximation \eqref{GetaUnrelaxed}. In the inset, the ratio between the two show a small deviation as temperature decreases. The shear viscosity shows a small sublinear deviation from $s/4\pi$ as $T$ decreases. }
\label{fig:plotGeta}
\end{figure}

The frequency dependence of  \eqref{hxyeom} will only generate $O(\omega^2)$ corrections which do not contribute to the retarded Green's function at leading order as $\omega\to0$.  So \eqref{hxyingoinglowomega} can be rewritten directly as
\begin{equation}
h_{2}(r)=h^{(reg)}_{2}(r)+i\omega\frac{s}{4\pi}\left(h^{(reg)}_{2}(r_h)\right)^2 h^{(sing)}_2(r)+O(\omega^2)\,.
\end{equation}
Close to the AdS boundary, $h_2^{(reg)}\sim1+k^2I_Y r^3/3+O(r^{4})$ while $h^{(sing)}_{2}\sim-r^3/3+O(r^{4})$, so that the full solution asymptotes to
\begin{equation}
h_{2}(r\to0)\sim1+\left(k^2I_Y-i\omega\frac{s}{4\pi}\left(h^{({reg})}_{2}(r_h)\right)^2\right)\frac{r^3}{3}+O(r^4,\omega^2)\,.
\end{equation}
From this equation, by applying standard holographic formul\ae\ relating the vev of an operator of dimension $3$ in $d=2$ to the asymptotic data of the dual field in the bulk \cite{Ammon:2015wua,Zaanen:2015oix,Hartnoll:2016apf}, we extract the low frequency limit of the shear retarded Green's function
\begin{equation}
\lim_{\omega\to0}G^R_{T^{xy}T^{xy}}(\omega,\vec{q}=0)=k^2I_Y-i\omega\frac{s}{4\pi}\left(h^{({reg})}_{2}(r_h)\right)^2\,.
\end{equation}

Matching to the hydrodynamic prediction \eqref{KuboUnrelaxed}, we obtain approximate, small $k$ expressions for the shear modulus and shear viscosity
\begin{equation}\label{GetaUnrelaxed}
G =k^2I_Y+O(k^4)\,,\quad \eta=\frac{s}{4\pi}\left(1-2k^2 \int_0^{r_h}dr_1\sqrt{\frac{B}D}\frac{1}{C}\int^{r_h}_{r_1}dr_2 \sqrt{BD}Y+O(k^4)\right).
\end{equation}
They match the exact numerical results very well, see figure \ref{fig:plotGeta}. We observe in figure \ref{fig:plotGeta} that $G$ decreases sharply above $T/\mu\gtrsim 0.1$: at high $T$, the spontaneous component of the system becomes very weak, but never vanishes.

\section{Diffusivity matrix from horizon data\label{section:unrelaxedDC}}

In this section, we compute the diffusivities $\sigma_o$, $\gamma_1$ and $\xi$  
in terms of {a combination of horizon and UV-sensitive data} through a DC analysis, drawing on previous 
works \cite{Donos:2014uba,Donos:2014cya,Amoretti:2017frz,Amoretti:2017axe,Donos:2018kkm,Gouteraux:2018wfe}. 
To that end, we turn on the following DC (linear in time) perturbation:
\begin{equation}
\label{pert1}
\delta a_x= a(r)-p_1(r)\, t\ ,\quad \delta h_{tx}= h_1(r)-p_2(r)\, t\ ,\quad \delta h_{rx}= h_3(r)\ ,\quad \delta\psi_x=\chi(r)-k\, t \delta v .
\end{equation}
The mode $\delta v$ corresponds to a vev, and in fact can be set to zero by a Galilean boost \cite{Donos:2018kkm,Gouteraux:2018wfe}. It is not fixed by the horizon analysis, and represents the freedom to slide the system back and forth. Upon turning on relaxation, this mode either becomes a source \cite{Amoretti:2018tzw} or is fixed by the horizon 
analysis \cite{Donos:2019tmo}.

All time dependence drops out from the linearized equations, provided
\begin{equation}
\label{pert2}
p_1(r)=p_1^{(0)}+\bar{E}\rho A\ ,\qquad p_2(r)=-\bar{E}\rho D \ ,
\end{equation}
where $p_1^{(0)}$ is a constant which will be fixed shortly {and $\bar E$ a source of the dual field theory to be defined more precisely below.}

The following boundary expansions are compatible with the equations of motion:
\begin{equation}
\label{pert3}
a(r)=a^{(1)}r+O(r^2)\ ,\quad h_1(r)=h_1^{(1)} r+O(r^2)\ ,\quad h_{3}(r)=O(r)\ ,\quad \chi(r)=\frac{\chi_{(0)}}{r}+\chi_{(1)} +O(r)\ ,
\end{equation}
provided we set 
\begin{equation}
\label{pert4}
p_1^{(0)}=-\bar{E}\left(\frac92 d_3+\rho\mu\right)+\frac{k}{\rho}(\phi_{(v)})^2\chi_0=\bar E\left(sT+k^2 I_Y\right)+\frac{\delta s_\varphi}{\rho}.
\end{equation}
This condition follows from requesting $\delta h_{rx}$ to fall off sufficiently fast in the UV. 
It differs from our previous work \cite{Amoretti:2017frz} by the term proportional to the phonon source $\delta s_\varphi=k(\phi_{(v)})^2\chi_{(0)}$.

Next, we define two bulk currents:
\begin{equation}
\label{ConsBulkCurrents}
\mathcal J(r)=\sqrt{-g}Z(\phi)F^{rx}=Z\sqrt{\frac{D}{B}}a'-\rho\frac{h_1}{C}\,,\quad \mathcal Q(r)=-\sqrt{\frac{D}{B}}h_1{}'+\frac{D' h_1}{\sqrt{BD}}-A\mathcal J\,,
\end{equation}
whose radial evolution is governed by the equations
\begin{equation}
\label{radialconseq}
\mathcal J'=0\,,\qquad \left(\mathcal Q+\delta v k^2\int_{0}^{r}d\tilde r\,\sqrt{DB}Y\right)'=0
\end{equation}
and which asymptote respectively to the boundary electric and heat currents: 
\begin{equation}
j=\mathcal J(0)=a^{(1)}\,,\qquad j_q\equiv T^{tx}-\mu j=\mathcal Q(0)=-3 h_{1}^{(1)}-\mu a^{(1)}\,.
\end{equation}
Using \eqref{radialconseq}, we can relate the boundary currents to data on the horizon:
\begin{equation}
\label{BdyCurrentFromHdata}
j=\mathcal J(r_h)\,,\qquad j_q=\mathcal Q(r_h)+k^2 I_Y\delta v\,.
\end{equation}

Horizon regularity of the solution in ingoing Eddington-Finkelstein coordinates $\textrm{v}=t+1/(4\pi T)\ln(r_h-r)+O(r_h-r)$ 
implies that the linear perturbation has the following near-horizon expansion:
\begin{equation}
\begin{split}
&a=\left[-\bar{E}\frac{s T+k^2 I_Y}{4\pi T}+\frac{\delta s_\varphi}{4\pi T \rho}\right]\log(r_h-r)(1+O(r_h-r))\ , \\
&h_1=-\bar{E} \frac{\rho I_Y}{ Y_h}-\frac{\delta s_\varphi}{k^2 Y_h}-\frac{s}{4\pi}\delta v+O(r_h-r)\ , \\
&h_3=-\frac{h_1}{4\pi T(r_h-r)}+O(1)\ .
\end{split}
\end{equation}

Evaluating the currents in \eqref{BdyCurrentFromHdata} at the horizon, we find
\begin{align}
\label{HorizonJ}
j&=\left[(s T+k^2 I_Y)Z_h+\frac{4\pi \rho^2 I_Y}{s k^2 Y_h}\right]\bar E+\left[\frac{4\pi \rho}{s k^2 Y_h}+\frac{Z_h}{\rho}\right]\delta s_\varphi
+\rho\delta v\,,\\
\label{HorizonQ}
j_q&=\frac{4\pi T \rho I_Y}{ Y_h}\bar E+\frac{4\pi T}{ k^2 Y_h}\delta s_\varphi+\left(s T+k^2 I_Y\right)\delta v\,.
\end{align}

In order to compute $\sigma_0$, $\gamma_1$ and $\xi$ we can use the relation between the hydrodynamic 
correlators \eqref{KuboUnrelaxedlambda}, \eqref{KuboUnrelaxedlambda2} and the expectations values of the related current densities, namely:
\begin{equation}
J_A=G_{AB}\,S_B\,,
\end{equation}
with $J_A=\{j,j_q,j_{\varphi} \}$, $S_A=\{E_x/(i \omega), -\nabla_x T/(i \omega T), \delta s_{\varphi} \} $ and where $j_\varphi=\dot\varphi$ (which can loosely be thought of as the contribution of the phonons to the electric current \cite{RevModPhys.60.1129}). 
The boundary electric field and temperature gradients are given by the boundary behaviour of the bulk fields:
\begin{equation}
E_x=-\lim_{r\to 0}\partial_t \left(\delta a_x(r,t)+\mu \delta h_{tx}(r,t)\right)\ ,\qquad \frac1T\nabla_x T=\lim_{r\to 0}\partial_t \delta h_{tx}(r,t)\,.
\end{equation}
{By rotating to the basis of currents $(j_{inc}\equiv\chi_{\pi j_q}j-\chi_{j\pi}j_q,\pi)$, $\bar E$ can be understood as the source of the incoherent current $j_{inc}$, \cite{Amoretti:2017frz}. The incoherent current is the part of the electric current which does not drag momentum, \cite{Davison:2015taa}. In other terms, $\chi_{j_{inc}\pi}=0$.}

 Using \eqref{pert2} and \eqref{pert4}, the electric field and temperature gradient can be expressed in terms of the sources $\bar{E}$ and $\delta s_{\phi}$:
\begin{equation}
E_x=(\chi_{\pi \pi}- \mu \rho)\bar{E}+\frac{\delta s_{\varphi}}{\rho} \ , \qquad \nabla_x T= \bar{E}\rho T \,,
\end{equation}
where $\chi_{\pi\pi}$ is given by \eqref{chiPP}.
We can now form the following linear combination of currents and relate them to the sources we have turned 
on using the hydrodynamic correlators \eqref{KuboUnrelaxedlambda}, \eqref{KuboUnrelaxedlambda2}:
\begin{align}
\label{CombinationCurrents1}
\chi_{\pi j_q}j-\chi_{j\pi}j_q&=\chi_{\pi\pi}^2\sigma_0\bar E+\chi_{\pi\pi}\left(\frac{\sigma_0}{\rho}-\gamma_1\right)\delta s_\varphi\,,\\
\label{CombinationCurrents2}
\mu j+j_q+\chi_{\pi\pi}j_\varphi&=-\bar E\gamma_1\chi_{\pi\pi}^2+\delta s_\varphi\chi_{\pi\pi}\left(X-\frac{\gamma_1}{\rho}\right)\,,
\end{align}
where $\chi_{\pi j_q}=\chi_{\pi\pi}-\mu\rho$. This follows from $j_q=\pi-\mu j$, as required by relativistic symmetry.
Undoing the term $\delta v$ in $\delta \psi_x$ with a Galilean boost amounts to shifting the currents $j$, $j_q$ by terms proportional to $\delta v$. These terms can in turn be absorbed by a redefinition of the velocity $v^x$, which finally gives a contribution to $j_\varphi=-\delta v$, recalling the Josephson relations \eqref{Josephson}.
Substituting \eqref{HorizonJ} and \eqref{HorizonQ} on the left hand side 
of \eqref{CombinationCurrents1} and \eqref{CombinationCurrents2}, we find that
\begin{eqnarray}
\label{diffDCsigmao}
\sigma_0&=&\frac{\left(s T+k^2 I_Y\right)^2}{\left(s T+\mu\rho+k^2 I_Y\right)^2}Z_h+\frac{4 \pi k^2 (I_Y)^2\rho^2}{s Y_h\left(s T+\mu\rho+k^2 I_Y\right)^2}\,,\\ \label{diffDCgamma1}
\gamma_1&=&-\frac{4\pi I_Y\rho\left(sT+\mu\rho\right)}{s Y_h\left(s T+\mu\rho+k^2 I_Y\right)^2}-\mu\frac{\left(s T+k^2 I_Y\right)}{\left(s T+\mu\rho+k^2 I_Y\right)^2}Z_h\,,\\\label{diffDCxi}
X&=&\frac{4\pi \left(sT+\mu\rho\right)^2}{k^2 s Y_h\left(s T+\mu\rho+k^2 I_Y\right)^2}+\frac{\mu^2 Z_h}{\left(s T+\mu\rho+k^2 I_Y\right)^2}\,,
\end{eqnarray}
where we have used \eqref{chiPP} for $\chi_{\pi\pi}$. {The dependence on UV-sensitive data is manifest through factors of $\mu$ and $I_Y$.\footnote{The common factor of $\chi_{\pi\pi}^2$ in the denominator can be removed by a choice of normalization.}}

As we already commented upon at the end of section \ref{section:acbdycorr}, the expressions \eqref{diffDCsigmao}-\eqref{diffDCxi} 
give an excellent match to the zero frequency limit of the ac correlators.

\section{Saturation of entropy bound and universal relaxation \label{section:bound}}

\begin{figure}[tbp]
\centering
\begin{tabular}{c}
\includegraphics[width=.8\textwidth]{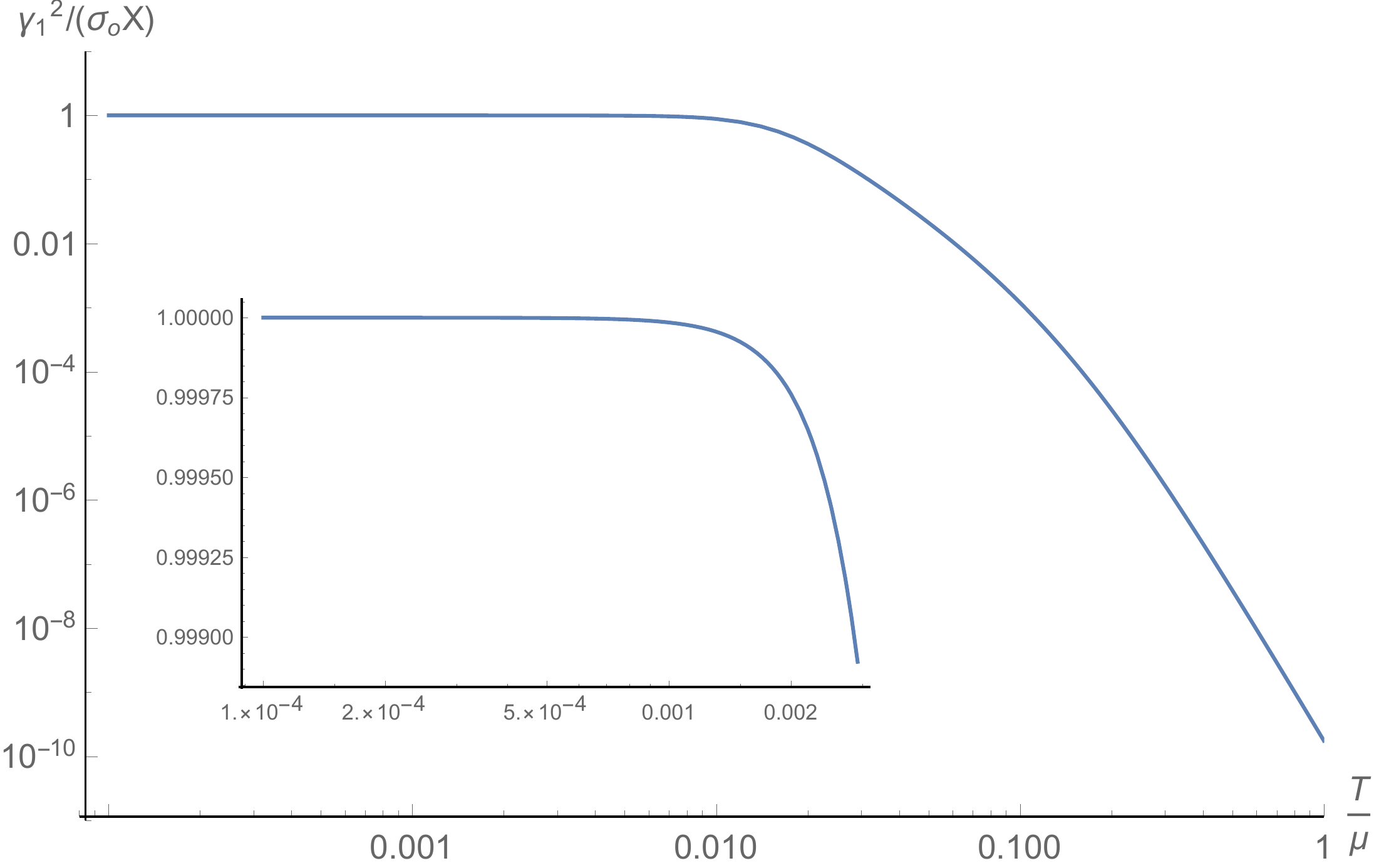}
\end{tabular}
\caption{At all temperatures, the entropy bound $\gamma_1^2\leq\sigma_o X$ holds. At low temperatures, the bound is saturated.}
\label{fig:entropybound}
\end{figure}

WC hydrodynamics predicts a bound coming from positivity of entropy production \cite{Delacretaz:2017zxd}, 
\begin{equation}
\label{entropybound}
\gamma_1^2\leq\sigma_o X\,,
\end{equation} 
which is obeyed by our DC expressions for the diffusivities:
\begin{equation}
\label{boundsaturation}
\sigma_0 X-\gamma_1^2=\frac{4\pi s T^2 Z_h}{k^2(\chi_{\pi\pi})^2Y_h}\geq0\,.
\end{equation}
See figure \ref{fig:entropybound}.

 Interestingly, for our specific choice of holographic model, the bound becomes saturated at low temperatures.  Indeed,  for the $z=+\infty$, $\theta=-\infty$, $-\theta/z=1$ family we have mostly focussed on, $Z_h\sim1/T$, $Y_h\sim T^0$ and $s\sim T$. Saturation of \eqref{entropybound} relates the diffusivities $\sigma_o$, $X$ and $\gamma_1$ to one another. This is reminiscent of the relation between second-order 
 transport coefficients of fluid hydrodynamics uncovered in \cite{Haack:2008xx}. Magnetohydrodynamics is another example where a bound originating from the positivity of entropy production 
 \cite{Grozdanov:2016tdf} is saturated in an explicit holographic realization \cite{Grozdanov:2017kyl}.

\begin{figure}[tbp]
\centering
\begin{tabular}{c}
\includegraphics[width=.8\textwidth]{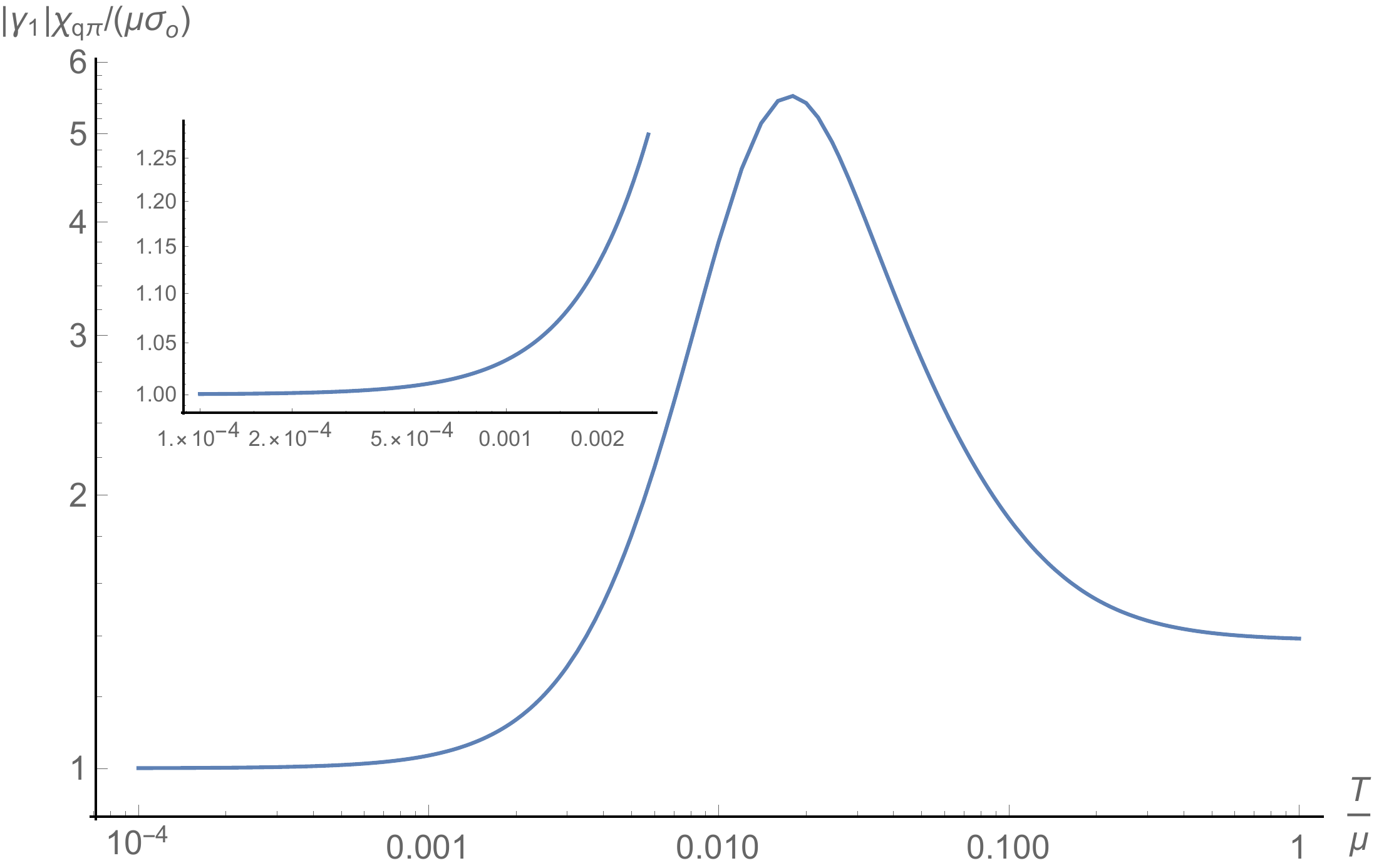}\\
\includegraphics[width=.8\textwidth]{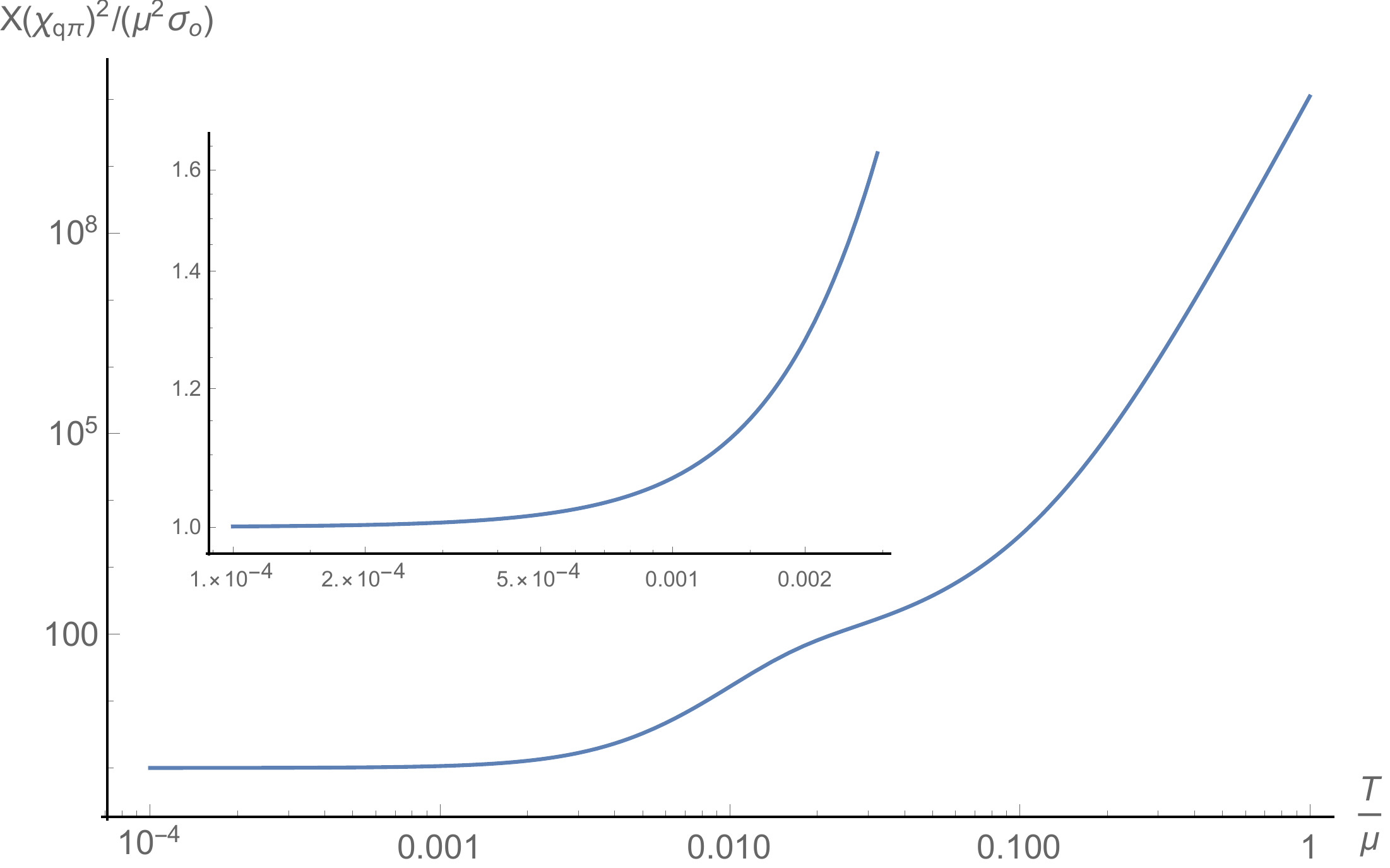}
\end{tabular}
\caption{At low temperatures, the entropy bound is saturated and the diffusivities $\gamma_1$, $X$ are controlled by the thermal diffusivity through \eqref{lowTdiff}.}
\label{fig:lowTdiff}
\end{figure}

To understand this feature, we consider the following Kubo formul\ae\ for the diffusivities:
\begin{eqnarray}
\label{Kubogamma1}
\gamma_1&=&-\lim_{\omega\to0}\frac1\omega\textrm{Im}G^R_{j\dot\varphi}(\omega,q=0)\,,\\\label{Kuboxi}
X&=&\lim_{\omega\to0}\frac1\omega\textrm{Im}G^R_{\dot\varphi\dot\varphi}(\omega,q=0)\,,\\
\sigma_o&=&\lim_{\omega\to0}\frac1\omega\textrm{Im}G^R_{jj}(\omega,q=0)\,.
\end{eqnarray}
The Kubo formul\ae\ \eqref{Kubogamma1}, \eqref{Kuboxi} involve the operator $\dot\varphi$, 
which suggests that a memory-matrix type analysis should apply \cite{Delacretaz:2019}. 
Relaxation of the Goldstones into the heat current provides a particularly appealing and universal mechanism for systems at finite temperature. 
It contributes to the low energy Hamiltonian via a term
\begin{equation}
\label{lowTjQrelax}
\Delta\mathcal H=\frac1{\chi_{\pi j_q}}\pi\cdot j_q\,.
\end{equation}
The full Hamiltonian will contain other contributions from non-hydrodynamic operators. Here we assume these other terms do not dominate at low temperatures.
The contribution from \eqref{lowTjQrelax} to $\dot\varphi$ is
\begin{equation}
\dot\varphi =i[\Delta\mathcal H,\varphi]= \frac{j_q}{\chi_{\pi j_q}}\,.
\end{equation}
{This equation shows that the time evolution of the Goldstone operator is governed by the heat current, as advertised.}
Plugging this in the Kubo formul\ae\ above relates the diffusivities $\gamma_1$, $\xi$ to the thermal diffusivity:
\begin{equation}
\label{lowTdiff}
\gamma_2=-\frac{\mu}{T}\gamma_1=\frac{\mu^2}{T\chi_{\pi j_q}}\sigma_o=\frac{\bar\kappa_o}{\chi_{\pi j_q}}\,,\qquad X=\left(\frac{\mu}{\chi_{\pi j_q}}\right)^2\sigma_o=\frac{T}{\chi_{\pi j_q}^2}\bar\kappa_o\,,
\end{equation}
where we have also used \eqref{RelHydro}.
For the $\theta=-\infty$, $z=+\infty$, $\theta/z=-1$ state we have considered in detail in this work, these expressions capture the correct low temperature behaviour of the diffusivities  (see figure \ref{fig:lowTdiff}), and lead to the saturation of the entropy bound \eqref{entropybound}.
At higher temperatures, other, non-universal relaxation channels will open. 
For the $\theta=-\infty$, $z=+\infty$, $\theta/z=-1$ holographic state, these non-hydrodynamic operators also contribute to the Hamiltonian \eqref{lowTjQrelax} 
but give subleading corrections as $T\to0$.\footnote{See \cite{Blake:2015epa} for another 
holographic example where momentum relaxation due to explicit translation symmetry breaking is controlled by the heat current.}

It is interesting to assess how generic the universal relaxation mechanism we have just described is. To do so, we consider the more general family of IR end points \eqref{skaska}. It is clear that if there is no running scalar $\phi$ and the IR is AdS$_2\times$R$^2$, the entropy bound is saturated since all factors in \eqref{boundsaturation} scale like $T^0$ at low temperatures, except of course for the explicit $T^2$ factor. When there is a running scalar, there are four classes of IR end points, characterized by the presence or absence of certain dangerously irrelevant deformations. These are discussed in some detail in \cite{Gouteraux:2014hca,Amoretti:2017frz,Amoretti:2017axe} and more recently in \cite{Davison:2018nxm}, so we will be brief here. 

There are two potentially dangerously irrelevant deformations, depending on whether the bulk fields dual to the conserved current and dual to the phonons support the deep IR solution or whether they decay faster than other fields in the bulk. These deformations can be usefully characterized by the scaling dimension of the coupling, which we denote by $\Delta_{A_o}$ and $\Delta_k$,  respectively. In order for the coupling to source an irrelevant/marginal deformation, $\Delta_{A_o,k}\leq0$. The four classes differ by whether both deformations are marginal, or only one of them is, or both of them are irrelevant. In the case we have studied numerically and for which the bound is saturated, both deformations are marginal $\Delta_{A_o}=\Delta_k=0$. {Their expressions in terms of the action couplings have been given in \eqref{irexponents}

Evaluating the right-hand side of \eqref{boundsaturation} on the solutions \eqref{skaska} together with the expressions \eqref{IRscalarcouplings}, \eqref{irexponents} for the IR couplings and recalling that $s\sim T^{(d-\theta)/z}$, the entropy bound is saturated provided
\begin{equation}
\frac2z\left(z-2+\Delta_{A_o}+\Delta_k\right)\geq0\,.
\end{equation}
We find that this condition is met provided
}
\begin{equation}
\label{parspacesaturation}
z>2 \quad \textrm{and}\quad \theta<2 \quad \textrm{and}\quad 2-z<\Delta_{A_o,k}\leq0\,.
\end{equation}
In particular, this implies the bound is never saturated for IR end points with relativistic symmetry $z=1$, which only happens when both deformations are irrelevant. If only one deformation is irrelevant, then the dimension of the coupling is bounded from below. If both are marginal $\Delta_{A_o,k}=0$, then the bound is saturated if the dynamical exponent $z$ is large enough, $z>2$.

{In the parameter space spanned by \eqref{parspacesaturation}, we observe that $(2+z-\theta)>-\Delta_{A_o,k}$. Recalling that $sT\sim T^{(2+z-\theta)/z}$ at low temperature, this suggests that $sT$ terms can be neglected at low temperatures in \eqref{diffDCsigmao}-\eqref{diffDCxi} against other terms that have a non-vanishing zero temperature limit, such as $\mu\rho$ or $k^2 I_Y$.\footnote{The irrelevant deformation $\Delta_{A_o,k}$ also source deviations from zero temperature. The inequality  $(2+z-\theta)>-\Delta_{A_o,k}$ implies that those sourced by $sT$ type terms are less important at low temperatures.} Doing so leads to 
\begin{eqnarray}
\label{diffDCsigmao2}
\sigma_0&=&\frac{\left(k^2 I_Y\right)^2}{\left(\mu\rho+k^2 I_Y\right)^2}\left(Z_h+\frac{4 \pi \rho^2}{s k^2Y_h}\right),\\
\gamma_1&=&-\frac{\mu k^2 I_Y}{\left(\mu\rho+k^2 I_Y\right)^2}\left(Z_h+\frac{4 \pi \rho^2}{s k^2Y_h}\right),\\\label{diffDCxi2}
X&=&\frac{\mu^2}{\left(\mu\rho+k^2 I_Y\right)^2}\left(Z_h+\frac{4\pi \rho^2}{k^2 s Y_h}\right).
\end{eqnarray}
We have verified that these expressions are consistent with our numerics at low temperature. It is straightforward to check that the expressions \eqref{diffDCsigmao2}-\eqref{diffDCxi2} saturate the entropy bound \eqref{entropybound}.

The prefactors are all temperature independent at low $T$, so all the low temperature dependence comes from the common factor in parenthesis. Since this common factor is the same for all three quantities, the saturation of the bound comes from the cancelation between the prefactors in the limit of low temperature, rather than from the vanishing of each individual term in \eqref{entropybound}.

On the other hand, requiring $(2+z-\theta)>-\Delta_{A_o,k}$ does not imply that the entropy bound is saturated. It would be interesting to further investigate these cases and to understand why the universal mechanism \eqref{lowTjQrelax} is not at play there.
}

\section{Transverse collective excitations at nonzero wavevector \label{sec:transfluc}}

\begin{figure}[tbp]
\centering
\begin{tabular}{cc}
\includegraphics[width=.58\textwidth]{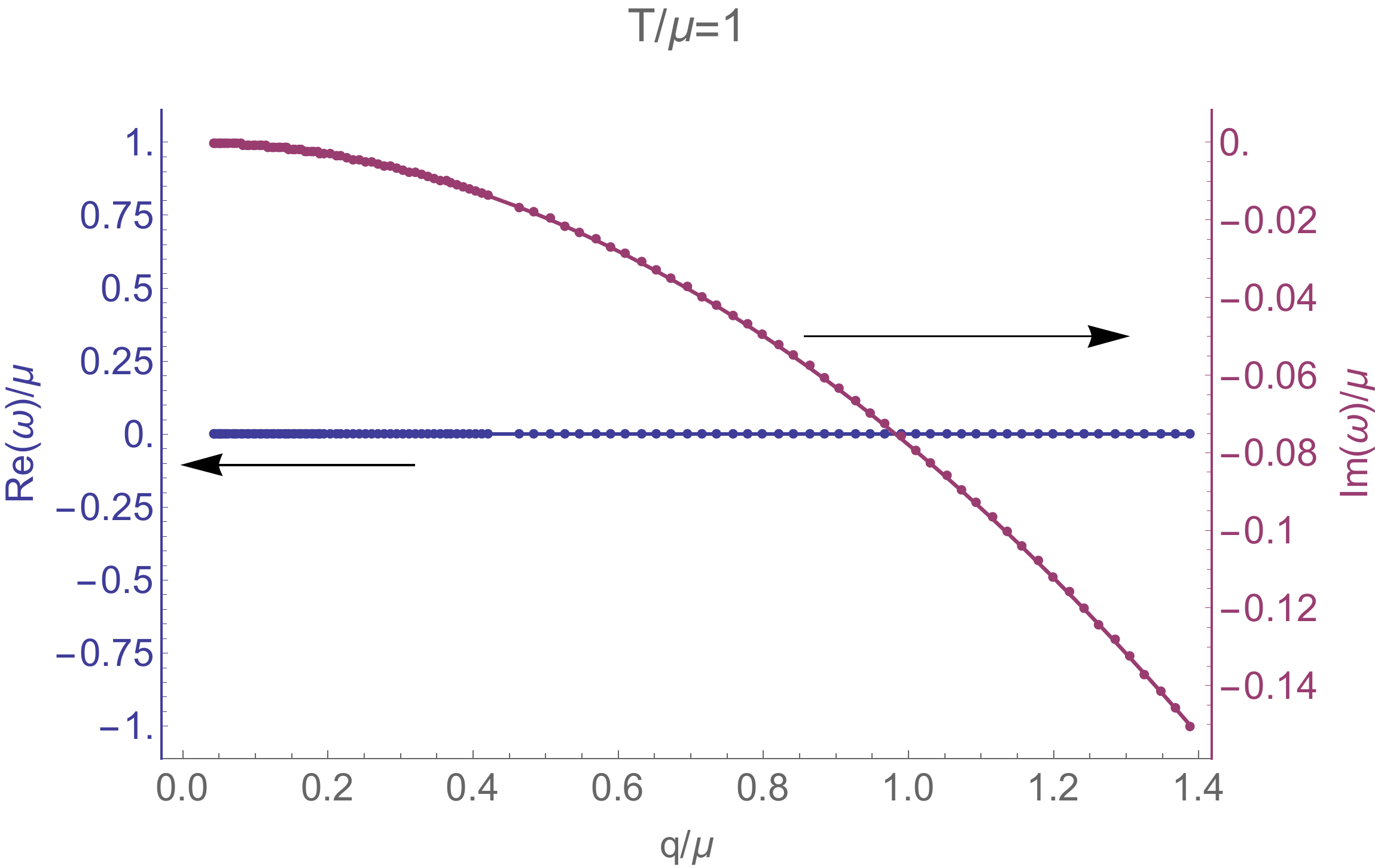}\\
\includegraphics[width=.58\textwidth]{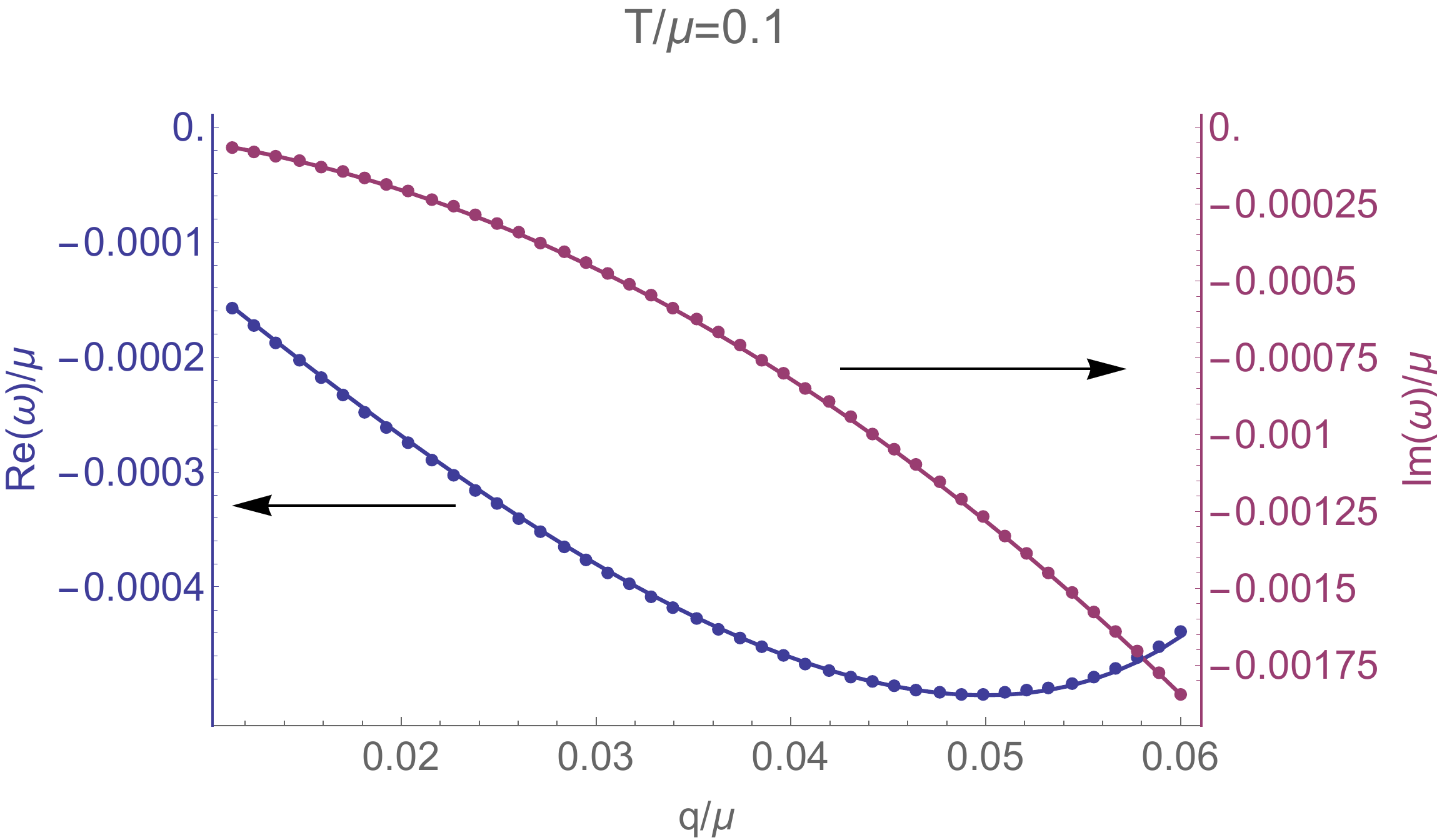}\\
\hfill\includegraphics[width=.58\textwidth]{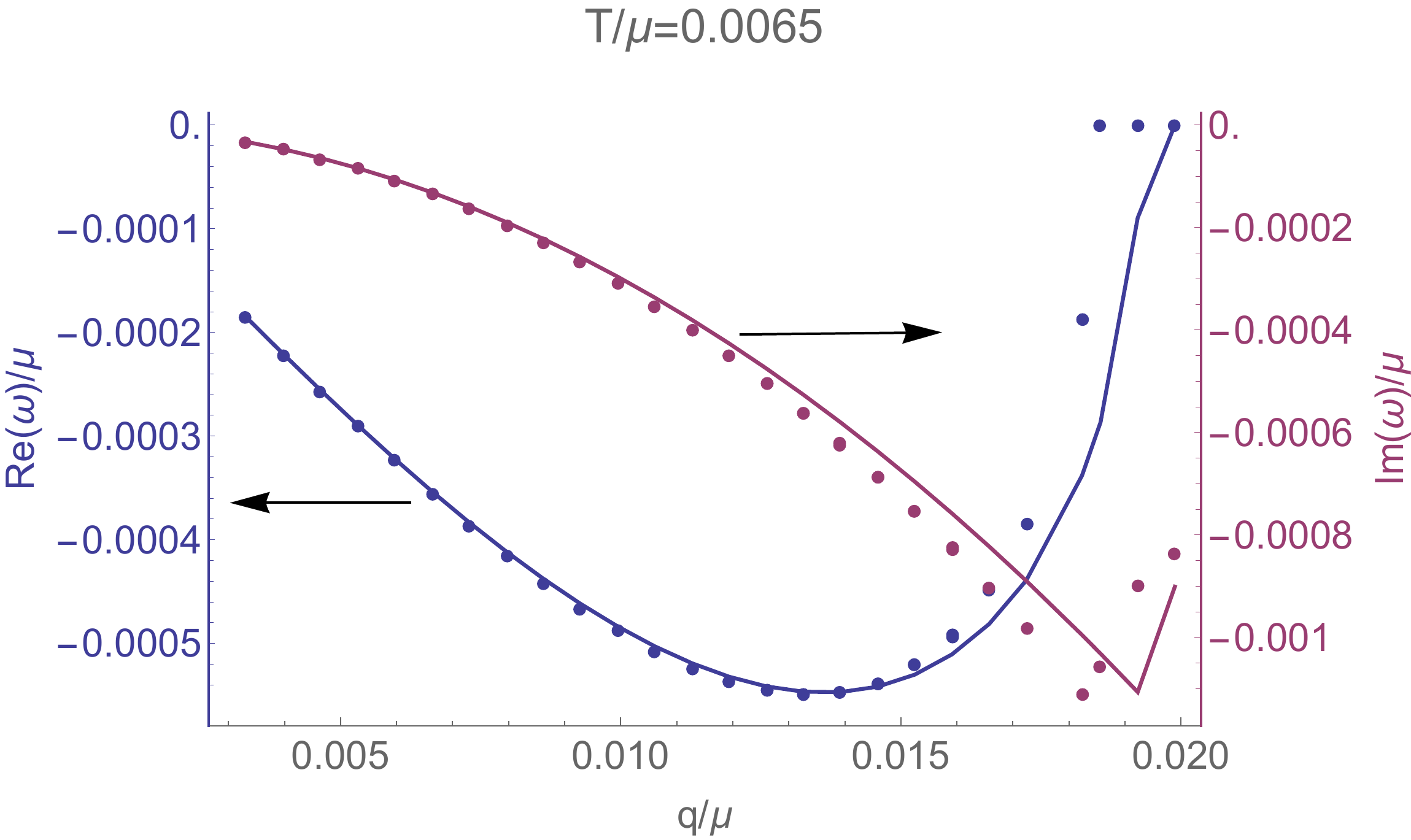}
\end{tabular}
\caption{Comparison of the exact location (dots, numerics) and hydrodynamic approximation \eqref{hydroapp} (solid lines) of the real (blue, left axis) and imaginary (magenta, right axis) part of the transverse hydrodynamic QNM at $T/\mu=1$ (top), $T/\mu=0.1$ (center) and $T/\mu=0.0065$ (bottom) vs $q/\mu$. In the top panel, the modes are purely imaginary, but would acquire a real part at smaller $q$ (not displayed).}
\label{fig:QNMvsq}
\end{figure}

WC hydrodynamics also predicts the existence of a pair of low energy modes in the transverse channel. They appear as zeroes of the denominator of the hydrodynamic retarded Green's functions and their dispersion relation is:\footnote{This dispersion relation is strictly speaking valid to order $q^2$ \eqref{hydroappgapless}, but keeping the expression \eqref{hydroapp} exact in $q$ derived from first order in gradients WC hydrodynamics leads to a better match to the numerics.}
\begin{equation}
\label{hydroapp}
\omega_{shear}=\frac12\left[-iq^2\left(\xi_\perp+\frac{\eta}{\chi_{\pi\pi}}\right)\pm q\sqrt{4\frac{G}{\chi_{\pi\pi}}-q^2\left(\frac{\eta}{\chi_{\pi\pi}}-\xi_\perp\right)^2}\right].
\end{equation}
At sufficiently low wavector $q$, these modes are gapless and propagate transverse sound waves:
\begin{equation}
\label{hydroappgapless}
\omega_{shear}=\pm\sqrt{\frac{G}{\chi_{\pi\pi}}}\,q-\frac{i}2 q^2\left(\xi_\perp+\frac{\eta}{\chi_{\pi\pi}}\right)+O(q^3)\,.
\end{equation}
On the other hand, for larger values of $q>2\sqrt{G\chi_{\pi\pi}}/(\eta-\xi_\perp\chi_{\pi\pi})$, the modes collide on the imaginary axis and are no longer propagating. Instead, they obey a pseudo-diffusive dispersion relation:
\begin{equation}
\label{hydroappgapped}
\omega_-=-i\xi_\perp q^2-\frac{i G}{\eta-\xi_\perp\chi_{\pi\pi}}+O(q^{-1})\,,\quad \omega_+=-i \frac{\eta}{\chi_{\pi\pi}}q^2+\frac{i G}{\eta-\xi_\perp\chi_{\pi\pi}}+O(q^{-1})\,.
\end{equation}
Observe that while the sign of the $q^0$ damping term is positive for one of the modes, this does not lead to an instability since $q$ is large. This collision is captured by the hydrodynamic dispersion relation, since the inequality on $q$ is true if $G$ is small enough. We expect $G$ to be small also close to a phase transition between a translationally ordered and a translationally disordered phase.

We compare the exact location of the quasinormal modes determined numerically to the hydrodynamic approximation in figure \ref{fig:QNMvsq} as a function of $q/\mu$, and find they agree very well, including at very low temperatures.  We typically observe deviations when $q\gtrsim T$, which suggests that temperature is the cut-off of the effective description in terms of WC hydrodynamics, whether $T$ is large or small compared to the chemical potential $\mu$. This gives strong evidence that the low energy excitations of our system are well described by WC hydrodynamics at all temperatures, and further confirms that our holographic model should be interpreted as describing the low energy dynamics of (pseudo)phonons coupled to conserved currents. We emphasize that the same dispersion relation is valid at both low and high temperature. Similar results on the low-temperature hydrodynamics of translation-invariant black holes have been reported in past literature  \cite{Edalati:2010hk,Edalati:2010pn,Davison:2013bxa}.

\begin{figure}[tbp]
\centering
\begin{tabular}{cc}
\includegraphics[width=.8\textwidth]{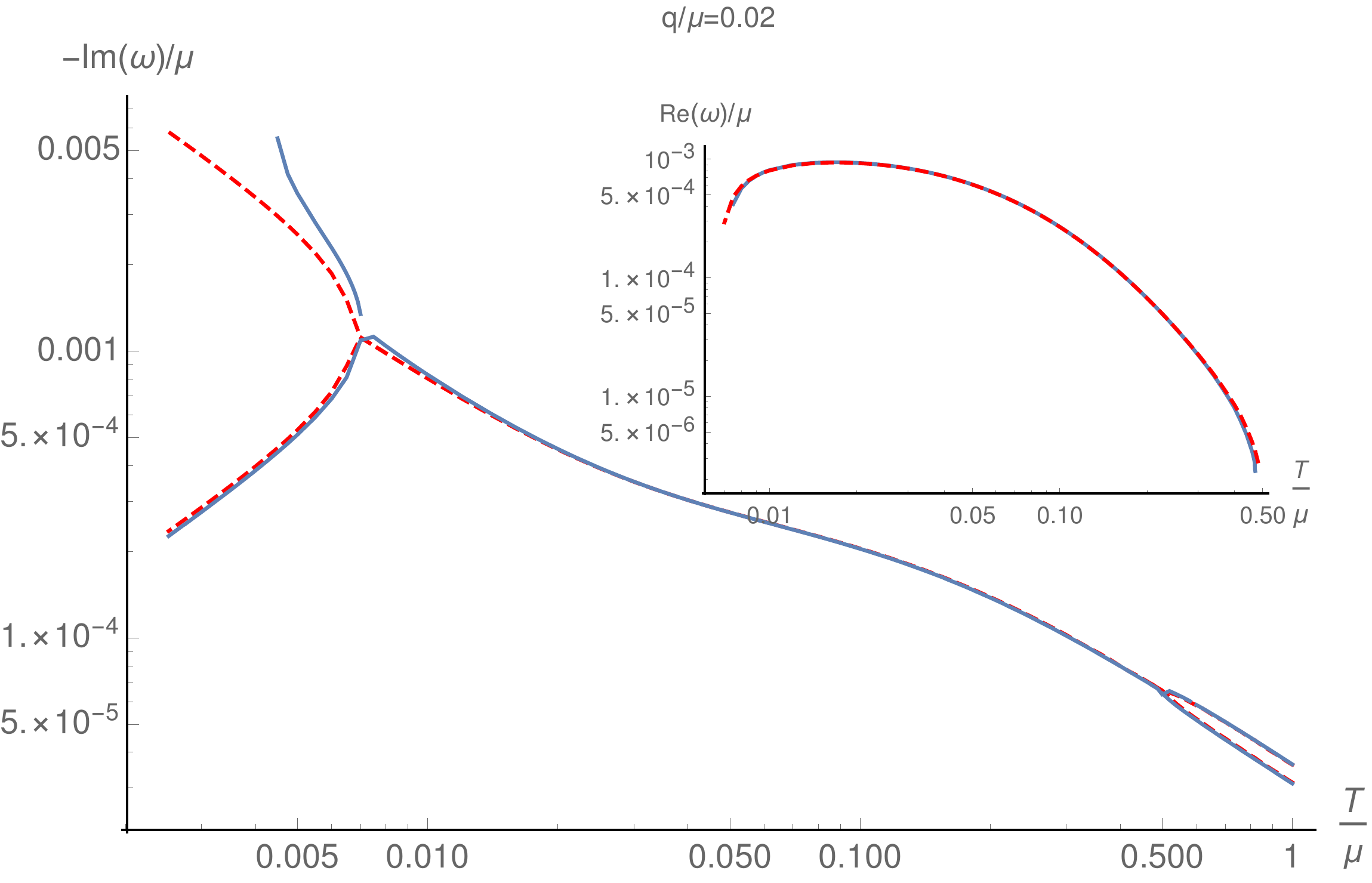}
\end{tabular}
\caption{Comparison of the exact location (numerics, solid blue) and hydrodynamic approximation \eqref{hydroapp} (dashed, red) of the real (inset) and imaginary part of the transverse hydrodynamic QNM at $q/\mu=0.02$ vs $T/\mu$. At intermediate temperatures, they follow very well the dispersion relation of shear sound waves predicted by WC hydrodynamics. At both very high and very low $T$, the modes collide and become purely imaginary and pseudo-diffusive. Keeping $T/\mu$ fixed, they modes acquire a nonzero real part as $q$ is lowered, see figure \ref{fig:QNMvsq}.}
\label{fig:QNMfiniteq}
\end{figure}

In figure \ref{fig:QNMfiniteq}, we display the QNMs location as a function of $T/\mu$ and compare it  to \eqref{hydroapp}. Increasing temperature, the shear sound modes collide on the imaginary axis. At these temperatures, the shear modulus of the system becomes very small (see figure \ref{fig:plotGeta}), and this is the underlying reason behind the high temperature pole collision (see figure \ref{fig:QNMfiniteq}). The same loss of shear rigidity would occur in a real crystal right before the phase transition to a liquid phase: the solid would no longer support gapless, transverse sound modes. The two high temperature branches are well-approximated by \eqref{hydroappgapped}, with $\omega_+$ giving the longest-lived mode. In other words, as the spontaneous component of the system is becoming very small, the system resembles more and more a fluid (consistently with the gap of \eqref{hydroappgapped} being proportional to $G$, which is going to zero as $T\to+\infty$).

\begin{figure}[tbp]
\centering
\begin{tabular}{cc}
\includegraphics[width=.8\textwidth]{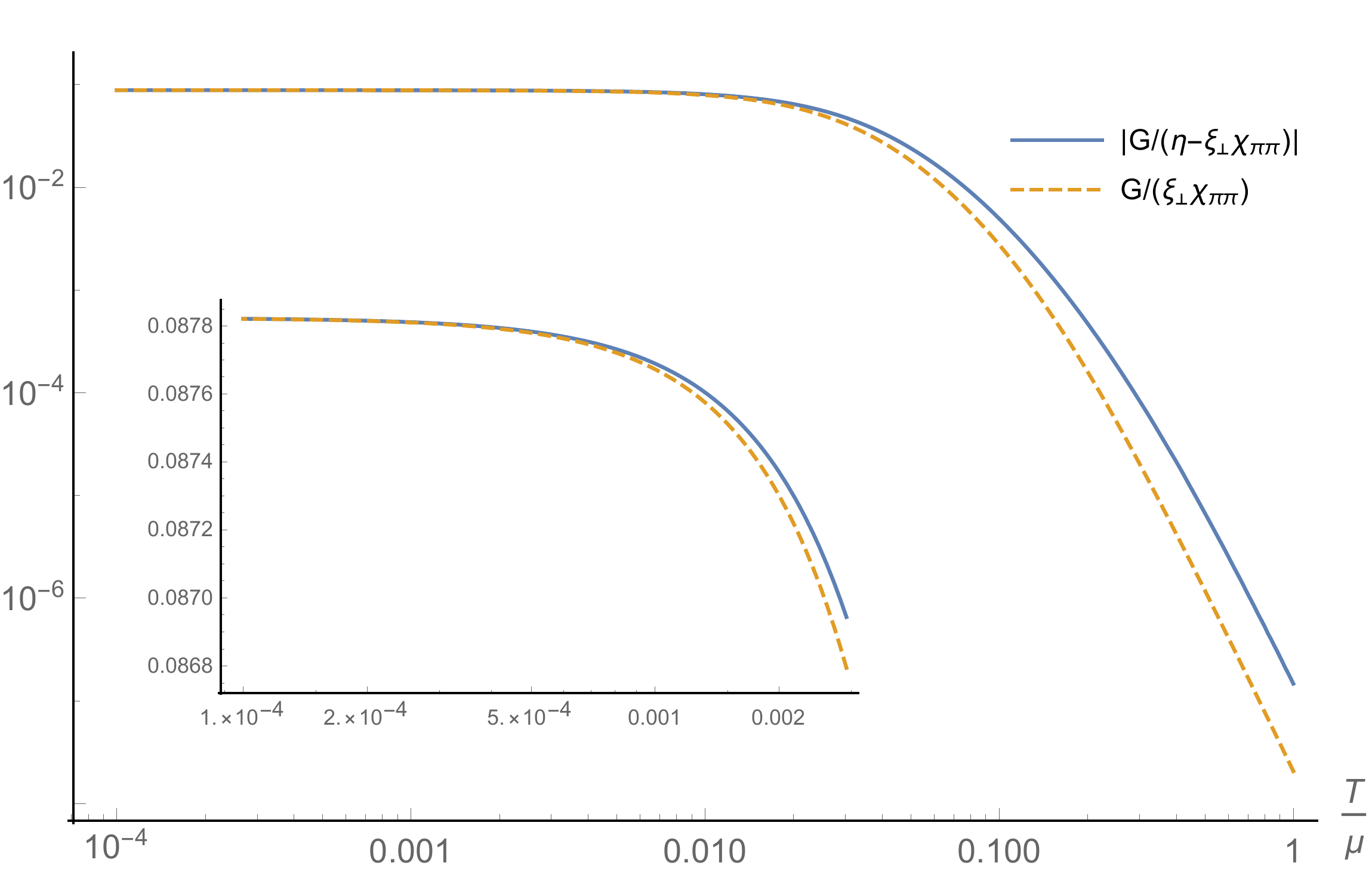}
\end{tabular}
\caption{At low temperatures, the gap at large wavevector is dominated by the phonon diffusivity $\xi_\perp$ rather than by the shear viscosity $\eta$.}
\label{fig:QNMfiniteqgap}
\end{figure}

At low temperatures, the QNMs also undergo a collision, which is now driven by the fact that $\xi_\perp\sim1/T$ is becoming very large. Once again, the longest-lived mode is well-approximated by $\omega_+$ in \eqref{hydroappgapped}. The gap is inversely proportional to $\xi_\perp$ and decreases with $T$, see figure \ref{fig:QNMfiniteqgap}. The system resembles more and more a fluid since the phonons relax faster (they have a larger diffusivity).

Similar collisions between high temperature, gapped, pseudo-diffusive modes leading to low temperature propagating modes have been reported in previous holographic literature \cite{Davison:2014lua,Grozdanov:2018ewh,Baggioli:2018vfc,Grozdanov:2018fic}. There are important differences. In our case, translations are broken spontaneously, not explicitly. As a consequence, at the longest distances $q\to0$, the modes are always gapless, sound modes. The collision we report is `coherent' or in other words, can be described by a hydrodynamic effective theory (WC hydrodynamics). In contrast, in \cite{Davison:2014lua,Baggioli:2018vfc,Grozdanov:2018fic} the breaking is always explicit and the effective theory is momentum-relaxed fluid hydrodynamics, and breaks down below a certain temperature where momentum is no longer approximately long-lived. In \cite{Grozdanov:2018ewh}, translations are broken spontaneously, but the transverse propagating modes are gapped. Gapless propagating modes are found for a nonzero density of defects, and it would be interesting if they became gapped at shorter distances, while remaining within the regime of validity of the hydrodynamic description.

Recently, \cite{Ammon:2019wci} have also reported similar collisions in a holographic massive gravity model.

\section{Outlook}

The main results we have obtained are as follows. Firstly, we have shown that all of the diffusivities that characterize the low energy dynamics of a holographic state breaking translations spontaneously can be computed in terms of {a combination of horizon and UV-sensitive data}. We have focussed on a specific model breaking translations homogeneously, and it would be natural to extend this computation to inhomogeneous, more realistic setups. {Moreover, in \cite{Donos:2018kkm,Amoretti:2017frz,Amoretti:2017axe,Gouteraux:2018wfe}, the incoherent conductivity was computed for thermodynamically stable phases and solely depends on horizon data}.\footnote{This statement is strictly true for the conductivity associated to the incoherent current defined as $j_{inc}=\chi_{\pi q}j-\chi_{\pi j}j_q$. Then, $\sigma_{inc}=\chi_{\pi\pi}^2\sigma_o$. For other choices of normalization, some overall factors of $\chi_{\pi\pi}$ appear in the incoherent conductivity.}

We have also shown how the low temperature behaviour of these diffusivities and the saturation of an entropy bound stemming from Wigner crystal hydrodynamics is explained by the universal relaxation of the phonons into the heat current. We have studied whether this depends on the infra-red end point controlling the system at zero temperature. We found that the entropy bound is not saturated in the presence of sufficiently irrelevant deformations, or for values of the dynamical exponent $1\leq z\leq2$. 

Finally, we have studied the spectrum of transverse collective excitations and discovered that for sufficiently low wavectors the pair of longest-lived modes are well captured by the dispersion relation following from WC hydrodynamics, including at low temperatures where the system is governed by the IR end point consistent with previous low temperature holographic studies \cite{Edalati:2010hk,Edalati:2010pn,Davison:2013bxa}. We observed deviations as $q\gtrsim T$. In particular, this means that the hydrodynamic description extends outside of the regime $q\ll T$ to $q\simeq T$ where gradients are of the same order as the thermal scale. This `unreasonable effectiveness' of hydrodynamics could perhaps be understood along the lines of the recent work \cite{Grozdanov:2019kge}.

We have only carried out explicit numerical calculations for the special case $z=-\theta=+\infty$. As we have observed, the saturation of the entropy bound will not occur for phases with $z\leq2$, in particular for $z=1$ in the presence of irrelevant deformations. It was shown in \cite{Davison:2018ofp,Davison:2018nxm} that in these cases new long-lived modes emerge at low temperature and lead to a breakdown of the hydrodynamic description at earlier times than that set by temperature. A natural future direction would be to study these cases with spontaneous translation symmetry breaking.

If translations are explicitly broken, the phonons obtain a mass $m$ and are damped, with a phase relaxation rate $\Omega$ (this phase relaxation rate is of a different microscopic origin than the phase relaxation rate sourced by the proliferation of topological defects \cite{PhysRevLett.41.121} but appears in the effective theory in exactly the same way). WC hydrodynamics in the presence of relaxation also predicts a bound on the relaxation parameters \cite{Delacretaz:2017zxd}
\begin{equation}
\label{relaxedbound}
\gamma_1^2\leq \textrm{min}\left(\sigma_oX,\frac{\sigma_o\Omega}{Gm^2}\right).
\end{equation}
In \cite{Amoretti:2018tzw}, we showed that turning on a small source for the scalar $\phi$ in our holographic model led to a phase whose low energy effective theory was relaxed WC hydrodynamics, until very low temperatures. The bound \eqref{relaxedbound} is also saturating as temperature decreases, suggesting that the pseudo-phonons also relax universally into the heat current. It would be interesting to understand how this depends on the infra-red end point, the relaxation mechanism and the details of how translations are broken in the holographic model. By now, many other models of pinned translational order are available \cite{Jokela:2017ltu,Alberte:2017cch,Andrade:2017cnc,Andrade:2017ghg,Andrade:2018,Donos:2019tmo,Ammon:2019wci}. 

In the presence of a magnetic field, the longitudinal and transverse phonons hybridize into a magnetophonon and a magnetoplasmon \cite{PhysRevB.18.6245}. For strong magnetic fields and in the presence of disorder, the magnetophonon is weakly gapped and a hydrodynamic theory can be formulated \cite{Delacretaz:2019}, which appears to fit well some of the data on two-dimensional electron systems. Universal relaxation of the magnetophonon into the electric current may dominate (compared to dislocation-mediated melting) in more disordered samples, and it would be interesting to study this holographically (see \cite{Romero-Bermudez:2019lzz} for a recent study of plasmonic response at zero field in the holographic model studied here).

\vskip1cm

\begin{acknowledgments}
We would like to thank Riccardo Argurio, Marco Fazzi, Saso Grozdanov, Javier Mas, Alfonso Ramallo and Javier Tarr\'\i o for stimulating and insightful discussions. BG would especially like to thank Luca Delacr\'etaz, Sean Hartnoll and Anna Karlsson for numerous insightful discussions on Wigner crystal hydrodynamics. BG has been partially supported during this work by the Marie Curie International Outgoing Fellowship nr 624054 within the 7th European Community Framework Programme FP7/2007-2013 and by the European Research Council (ERC) under the European Union’s Horizon 2020 research and innovation programme (grant agreements No 341222 and No 758759).  DM has been funded by the Spanish grants FPA2014-52218-P and FPA2017-84436-P by Xunta de Galicia GRC2013-024, by FEDER and by the María de Maeztu Unit of Excellence MDM-2016-0692. D.A. is supported by the `Atracci\'on del Talento' programme (Comunidad de Madrid) under grant 2017-T1/TIC-5258 and by Severo Ochoa Programme grant SEV-2016-0597 and FPA2015-65480-P (MINECO/FEDER). D.A. and D.M. thank the FRont Of pro-Galician Scientists for unconditional support. 
\end{acknowledgments}

\appendix

\section{AC boundary correlators at zero density \label{sec:bdycorrzerodensitysp}}

First, we need to figure out how to extract the retarded Green's functions \eqref{KuboUnrelaxedlambda2} from the asymptotic data. The relevant part of the renormalized on-shell action at quadratic order in the fluctuations \eqref{renspontaneous} (keeping only $h_{1,(0)}$ and $\delta \psi_{(-1)}$ nonzero) is
\begin{equation}
S_{\rm ren}^{(2)}=\left.\int d\omega \left[-\frac32\delta h_{1,(0)}\delta h_{1,(3)}+\frac32d_{(3)}(\delta h_{1,(0)})^2+(\phi_{(v)})^2\delta \psi_{(-1)}\delta \psi_{(0)}\right]\right|_{\delta a_{(0)},\delta h_{2,(0)}=0}
\end{equation}
so that the retarded Green's function are
\begin{equation}
\label{GRTtyTty}
\begin{split}
G^R_{T^{tx}T^{tx}}(\omega,q=0)=\frac{3\delta h_{1,(3)}}{\delta h_{1,(0)}}\,,&\quad G^R_{T^{tx}\delta \psi_x}(\omega,q=0)=\frac{3\delta h_{1,(3)}}{\delta \psi_{(-1)}}\,,
\\
 G^R_{\delta \psi_xT^{tx}}(\omega,q=0)=(\phi_{(v)})^2\frac{\delta \psi_{(0)}}{\delta h_{1,(0)}}\,,\quad& G^R_{\delta \psi_x\delta \psi_x}(\omega,q=0)=(\phi_{(v)})^2\frac{\delta \psi_{(0)}}{\delta\psi_{(-1)}}\,.
\end{split}
\end{equation}

We start by explicitly identifying the source of the boundary phonon. This is done by plugging the UV expansions \eqref{flucUVexpsp} for the bulk fluctuations and solving the equations of motion, which give the following relation
\begin{equation}
\delta h_{1,(3)}=\frac{i}{\omega}\frac{k(\phi_{(v)})^2\delta\psi_{(-1)}}{3}
\end{equation}
from which we immediately deduce
\begin{equation}
 G^R_{T^{tx}\delta \psi_x}(\omega,q=0)=\frac{3\delta h_{1,(3)}}{\delta \psi_{(-1)}}=k(\phi_{(v)})^2\frac{i}{\omega}\,.
\end{equation}
In order for this expression to match the off-diagonal Green's function $G^R_{\pi \varphi}$ in \eqref{KuboUnrelaxedlambda2}, we see that we should identify the phonon source as $\delta s=k(\phi_{(v)})^2\delta \psi_{(-1)}$.

To continue, we now need to solve the equations of motion away from the boundary.
It is helpful to define the variable \cite{Davison:2018nxm}
\begin{equation}
\Pi_x\equiv-\frac{{h^x_t}'+i\omega h^x_r}{\left(D/C\right)'}\,,
\end{equation}
which obeys the decoupled equation of motion
\begin{equation}
\frac{d}{dr}\left[\sqrt{\frac{D}{B}}\frac{1}{YC}\left(\frac{C^2}{\sqrt{BD}}\left(\frac{D}{C}\right)'\right)^2\Pi_x'\right]+\omega^2\sqrt{\frac{B}{D}}\frac{1}{YC^2}\left(\frac{C^{d/2+1}}{\sqrt{BD}}\left(\frac{D}{C}\right)'\right)^2\Pi_x=0.
\end{equation}
Then we can expand $\Pi_x$ in $\omega$
\begin{equation} 
\label{Piexp}
\Pi_x(r)=\left(s T+k^2\int^{r_h}_{r}\sqrt{BD}Y\right)\left(1-\frac{r}{r_h}\right)^{-i\omega/(4\pi T)}\left(\pi_o+\frac{i\omega}{4\pi T}\pi_1(r)+\dots\right)
\end{equation}
and solve order by order, imposing ingoing boundary conditions. At leading order, $\pi_o$ is just a constant. The solution for $\pi_1$ which is regular at the horizon is
\begin{equation}
\label{solPi1}
\pi_1(r)=i \pi_o\int_{r_h}^r dr_1\left[\frac{1}{r_1-r_h}+(4\pi T)^3\frac{C_h}{Y_h}\frac{\sqrt{B} C Y}{\sqrt{D}\left(s T+k^2\int^{r_h}_{r_1}\sqrt{BD}Y\right)^2}\right].
\end{equation}
To go through these manipulations, it is helpful to keep in mind the relation \eqref{radial1}:
\begin{equation}
\label{radialbulk}
\frac{C^{2}}{\sqrt{BD}}\left(\frac{D}{C}\right)'=-sT-k^2\int^{r_h}_{r}\sqrt{BD}Y.
\end{equation}
The boundary expansion of the $\Pi_x$ variable is
\begin{equation}
\Pi_x=\frac{i}{\omega}k(\phi_{(v)})^2\delta\psi_{(-1)}+\frac{i\omega}{2}k(\phi_{(v)})^2\delta\psi_{(-1)}r^2+\frac13k(\phi_{(v)})^2\left(k\delta h_{1,(0)}+i\omega\delta\psi_{(0)}\right)r^3+\dots
\end{equation}
To establish this, it is necessary to push the boundary expansions to higher order in $r$ than written in \eqref{flucUVexpsp}. Also, we note that the combination that appears at $O(r^3)$ is gauge invariant. So from \eqref{GRTtyTty} the retarded Green's function for $\delta\psi_x$ is
\begin{equation}
G^R_{\delta\psi_x\delta\psi_x}(\omega,q=0)=\frac{(\phi_{(v)})^2}{2\omega^2}\lim_{r\to0}\frac{\Pi_x^{(3)}(r)}{\Pi_x(r)}\,.
\end{equation}
Putting together \eqref{Piexp} and \eqref{solPi1}, we obtain after expanding at low frequency
\begin{equation}
\label{Gpsipsizerodensity}
G^R_{\delta\psi_x\delta\psi_x}(\omega,q=0)=\left(k(\phi_{(v)})^2\right)\left[\frac1{\left(\frac92d_{(3)}\right)\omega^2}+\frac{4\pi s T^2}{k^2 Y_h\left(\frac92d_{(3)}\right)^2}\frac{i}{\omega}\right]
\end{equation}
The momentum static susceptibility was computed in \cite{Amoretti:2017frz}
\begin{equation}
\chi_{\pi \pi}= -\frac92d_{(3)}=sT+k^2\int_{r_h}^0 \sqrt{BD}Y\,.
\end{equation}
Observe that our boundary expansions imply that the integral over the whole bulk does not diverge at the boundary. Plugging into \eqref{Gpsipsizerodensity} and comparing with the phonon retarded Green's function $G^R_{\varphi\varphi}$ in \eqref{KuboUnrelaxedlambda2} leads to identifying the phonon as
\begin{equation}
\varphi_x=\frac{\delta\psi_{(0)}}{k(\phi_{(v)})^2}
\end{equation}
and the ratio of the transverse crystal diffusivity to the shear modulus as
\begin{equation}
\label{AppXiovGzerodensity}
\frac{\xi_\perp}{G}=\frac{4\pi s T^2}{k^2 Y_h\left(sT+k^2\int_0^{r_h}  \sqrt{BD}Y\right)^2}\,.
\end{equation}
This result matches the zero density limit of \eqref{diffDCxi}.

\section{Numerics }\label{app:numerics}

In this appendix we describe how we construct the numerical solutions
of the holographic model \eqref{action} relevant for the analysis presented
in the main text.

\subsection{Black hole geometries}

The action \eqref{action} admits black hole solutions asymptotic to AdS
which realize holographically the (pseudo-)spontaneous breaking of translations.
In order to find those geometries
we take the following Ansatz for the metric and matter fields 
\begin{align}
&ds^2={1\over r^2}\left(-u(r)dt^2+{1\over u(r)}dr^2+c(r)(dx^2+dy^2)\right)\,,
\label{eq:metric}\\
&A=A_t(r)dt\,,\quad \phi=\phi(r)\,,\quad \psi_I=k x^I\,,\quad x^I=\{x,y\}.
\end{align}

The resulting equations of motion can be reduced to a system of four ordinary differential equations (three are second order and one is first order). 
For the potentials in eq. \eqref{eq:potentials} it is easy to find the following UV 
asymptotic solution:
\begin{subequations}
	\begin{align}
	&\phi(r)=\lambda\,r+v\,r^2+\lambda\left(k^2+{7\over36}\lambda^2\right)\,r^3+O(r^4)\,,\label{eq:phiuv}\\
	&A_t(r)=\mu-\rho\,r-{\lambda\,\rho\over 2\sqrt{3}}r^2
	-{\rho\over36}\left(5\lambda^2+4\sqrt{3}\,v\right)\,r^3+O(r^4)\,,\\
	&u(r)=1-{\lambda^2\over4}\,r^2+u_3\,r^3+O(r^4)\,,\\
	&c(r)=1-{\lambda^2\over4}\,r^2-{1\over3}\lambda\,v\,r^3+O(r^4)\,,
	\end{align}
	\label{eq:uvback}
\end{subequations}
where higher order coefficients are functions of $\lambda$, $v$, $\rho$,
and $u_3$.
This is not the most general asymptotic solution, but that with AdS asymptotics.

In the IR one can find the following near-horizon solution
\begin{subequations}
	\begin{align}
	&\phi(r)=\phi_{h}+O(r_h-r)\,,\qquad\qquad
	A_t(r)=A_{h,1}(r_h-r)+O((r_h-r)^2)\,,\\
	&u(r)=u_{h,1}(r_h-r)+O((r_h-r)^3)\,,\quad
	c(r)=c_h+c_{h,1}(r_h-r)+O((r_h-r)^2)\,,
	\end{align}
	\label{eq:irback}
\end{subequations}
where, for the potentials \eqref{eq:potentials}
\begin{equation}
u_{h,1}={c_h\,e^{-{\phi_h\over\sqrt{3}}}
	\left[6+e^{2\phi_h\over\sqrt{3}}(6-r_h^4\,A_{h,1})
	\right]
	-2k^2r_h^2\left(1-e^{\phi_h}\right)^2
	\over 2r_h\left(2c_h+c_{h,1}\,r_h\right)}
\end{equation}
determines the temperature of the black hole $T=-u_{h,1}/(4\pi)$, and further higher
order coefficients in \eqref{eq:irback} are also determined in terms of $\phi_h$,
$A_{h,1}$, $c_h$, and $c_{h,1}$.

It is easy to check that the equations of motion enjoy the scale invariance
$(t,x,y,r)\to\alpha\,(t,x,y,r)$, $A_t\to A_t/\alpha$, $k\to k/\alpha$
which we use to set the horizon radius $r_h=1$ in our numerical computations.
Next we generate numerical solutions by integrating the equations of motion from
the IR ($r=1$) to the UV ($r=0$).
However, a generic solution obtained this way will  not have the UV asymptotics
\eqref{eq:uvback}. In particular $c(r) = c_0 + c_1\, r+\dots$, and one would need to shoot
for $c_0 = 1$, and $c_1=0$ to get the AdS asymptotics \eqref{eq:uvback}. But one can make use of a further invariance of the equations under $(x,y)\to \beta\,(x,y)$, $k\to k/\beta$,
$c\to c/\beta^2$ to get $c_0 = 1$. This in practice means that $c_h$ is fixed and we are thus left with three IR parameters $\phi_h$, $A_{h,1}$, $c_{h,1}$; and one UV condition: $c_1=0$.
Therefore, we expect to obtain a two-parameter family of solutions.
We can choose those parameters to be the dimensionless ratios $T/\mu$
and $\lambda/\mu$.

In this work we are interested in solutions breaking translations spontaneously.
These are geometries where $\lambda/\mu$ vanishes but $v/\mu^2$ does not 
(solutions with a nontrivial $\phi$ that behaves asymptotically as $\phi\sim v\,r^2+\dots$).
As was shown in \cite{Amoretti:2017frz}, for the choice of potentials \eqref{eq:potentials} these solutions exist for any value of $T/\mu$.
In the following we will only consider geometries with $\lambda/\mu=0$.

\subsection{AC Fluctuations \label{subapp:flucs}}

In order to compute the boundary correlators analyzed in section \ref{section:acbdycorr} we consider the following consistent set of fluctuations
\begin{align}
\delta g_{tx}=h(r)\,e^{-i\omega t}\,,\quad \delta A_x=a(r)\,e^{-i\omega t}\,,\quad
\delta \psi_x=
\xi(r)\,e^{-i\omega t}\, .
\label{eq:ACflucs}
\end{align}
It is easy to check that at linear order in the fluctuations, the equations of motion
for $a$, $h$, and $\xi$ are a consistent set formed by two second order and one
first order differential equation.

For a spontaneous background geometry (where $\lambda=0$)
the asymptotic UV solution for the fluctuations at hand takes the form
\begin{subequations}
	\begin{align}
	&h(r)=r^{-2}\left(h_0+O(r^3)\right)\,, \\ 
	&a(r)=a_0+a_1\,r+O(r^2)\,, \\
	&\xi(r)=\xi_{-1}/r+\xi_0+ O(r)\,,
	\end{align}
	\label{eq:flucUV}
\end{subequations}
where higher order coefficients are functions of $h_0$, $a_0$, $a_1$, $\xi_{-1}$
and $\xi_0$. We shall also write down here the solution for the order $r^3$ contribution
to $h(r)$ since it will be used below in the computation of the holographic Green functions:
\begin{equation}
h_3={\rho\over3}\,a_0+{ik\over3\omega}\,v^2\,\xi_{-1}\,.
\label{eq:h3sol}
\end{equation}

We are interested in computing retarded propagators, hence we must require the
solutions to be ingoing towards the black hole horizon. 
They take the form
\begin{subequations}
	\begin{align}
	&h(r)\,e^{i{\omega\over u_{h,1}}}=h_{h,1}\,(r_h-r)+O((r_h-r)^2)\,, \\ 
	&a(r)\,e^{i{\omega\over u_{h,1}}}=a_h+O(r_h-r)\,, \\
	&\xi(r)\,e^{i{\omega\over u_{h,1}}}=\xi_h+O(r_h-r)\,,
	\end{align}
	\label{eq:flucIR}
\end{subequations}
with
\begin{equation}
h_{h,1}={a_h\,A_{h,1}\,e^{-{\phi_h\over\sqrt{3}}}\,r_h^2-\left(
	e^{\phi_h}-1\right)^2 \xi_h\,k/r_h
	\over i\omega/u_{h,1}-1}\,,
\end{equation}
and other higher order coefficients determined as well in terms of $a_h$, and $\xi_h$.

As it will be useful below, let us also point out that the following field configuration
\begin{equation}
h(r)=-i\omega\,{c(r)\over r^2}\,,\quad
a(r)=0\,, \quad \xi(r)=k\,,
\label{eq:puregaugefluc}
\end{equation}
solves the equations of motion, since it results from a diffeomorphism transformation
of the trivial solution.

\subsubsection{Computing the correlators}

We follow \cite{Kaminski:2009dh} and compute the holographic Green function $G^R_{AB}$ 
where $A,B =h,a,\xi$, as
\begin{equation}
G^R_{AB}={\cal B}+{\cal A}\,.\,V\,.\,S^{-1}\,.
\end{equation}
The matrices ${\cal A}$ and ${\cal B}$ can be read from the quadratic action
for the fluctuations as
\begin{equation}
S_{os}^{(2)}=\int d\omega \left(A_{IJ}\,{\mathfrak v}^J\,{\mathfrak s}^I+
B_{IJ}\,{\mathfrak s}^J\,{\mathfrak s}^I\right) ,
\end{equation}
where ${\mathfrak s}$ and ${\mathfrak v}$ are vectors made of the leading (sources)
and subleading (vevs) coefficients of the bulk fields that in the case at hand read
\begin{equation}
{\mathfrak s}=(h_0,a_0,\xi_{-1})\,,\quad
{\mathfrak v}=(h_3,a_1,\xi_0)\,.
\end{equation}
Therefore, rewriting the quadratic on-shell action \eqref{renspontaneous} in the gauge
(\ref{eq:metric},\ref{eq:ACflucs}) one gets
\begin{equation}
{\cal A}=\left(
\begin{array}{ccc}
-3 & 0 & 0 \\
0 & 1 & 0 \\
0 & 0 & v^2
\end{array}
\right)\,,\quad
{\cal B}=\left(
\begin{array}{ccc}
-u_3 & 0 & 0 \\
-\rho & 0 & 0 \\
0 & 0 & 0
\end{array}
\right)\,.
\end{equation}
Finally, $S$ and $V$ are respectively the matrices of sources and vevs constructed out of three independent solutions for the fluctuations 
$S=({\mathfrak s}^{I},{\mathfrak s}^{II},{\mathfrak s}^{III})$,
$V=({\mathfrak v}^{I},{\mathfrak v}^{II},{\mathfrak v}^{III})$. They read
\begin{equation}
S=\left(
\begin{array}{ccc}
h_0^{(I)} & h_0^{(II)} & -i\omega \\
a_0^{(I)} & a_0^{(II)} & 0 \\
\xi^{(I)}_{-1} & \xi^{(II)}_{-1} & 0
\end{array}
\right)\,,\quad
V=\left(
\begin{array}{ccc}
h_3^{(I)} & h_3^{(II)} & 0 \\
a_1^{(I)} & a_1^{(II)} & 0\\
\xi^{(I)}_0 & \xi^{(II)}_0 & k
\end{array}
\right)\,,
\end{equation}
in terms of the asymptotic coefficients \eqref{eq:flucUV}.
Notice that in
eq. \eqref{eq:flucIR} there are only two free IR parameters
($a_h$, $\xi_h$)
which allow us to construct two independent numerical solutions by shooting from the IR.
A third independent solution is given by \eqref{eq:puregaugefluc} and determines
the third column in both matrices above.

\subsection{Transverse fluctuations\label{subapp:k0flucs}}

In section \ref{sec:transfluc} we studied the transverse collective excitations of the
system at nonzero wavevector. In order to compute numerically
the corresponding QNMs we need to study the set of fluctuations
\begin{align}
\delta g_{tx}=h(r)\,e^{-i\omega t+q y}\,,\quad 
\delta g_{xy}=g(r)\,e^{-i\omega t+q y}\,,\quad
\delta A_x=a(r)\,e^{-i\omega t+q y}\,,\quad
\delta \psi_x=
\xi(r)\,e^{-i\omega t+q y}\, .
\end{align}
It is easy to check that at linear order in these fluctuations, the equations of motion
are a consistent set of one first order and three second order ordinary differential equations.
In a black hole background with $\lambda=0$ the asymptotic solutions for $h(r)$, $a(r)$,
and $\xi(r)$ have the same form as in eq. \eqref{eq:flucUV} above, while for $g(r)$ one
gets
\begin{equation}
g(r)=r^{-2}\left(g_0+O(r^2)\right)\,.
\label{eq:fluctransUV}
\end{equation}

Towards the black hole horizon the solution for the ingoing fluctuations also takes the same form as in Sec. \ref{subapp:flucs}, namely eq. \eqref{eq:flucIR}, with the addition of
\begin{equation}
g(r)\,e^{i{\omega\over u_{h,1}}}=g_{h}+O(r_h-r)\,.
\end{equation}

Now, to compute the QNMs in this sector we will follow \cite{Kaminski:2009dh} and employ
the so-called determinant method. Hence we need to obtain four independent solutions
for the fluctuations, and construct the following matrix of sources
\begin{equation}
S=\left(
\begin{array}{cccc}
h_0^{(I)} & h_0^{(II)} & h_0^{(III)} & -i\omega \\
g_0^{(I)} & g_0^{(II)} & g_0^{(III)} & iq \\
a_0^{(I)} & a_0^{(II)} & a_0^{(III)} & 0\\
\xi^{(I)}_{-1} & \xi^{(II)}_{-1} & \xi^{(III)}_{-1} & 0
\end{array}
\right)\,,
\label{eq:sourcematrixtrans}
\end{equation}
where in order to generate the fourth column we have used the pure gauge solution
\begin{equation}
h(r)=-i\omega\,{c(r)\over r^2}\,,\quad
g(r)=iq\,{c(r)\over r^2}\,,\quad
a(r)=0\,, \quad \xi(r)=k\,.
\label{eq:puregaugefluctrans}
\end{equation}
Finally, the QNMs, namely the complex frequencies where the holographic Green functions have a pole, are given by the values of $\omega$ for which the determinant of the
matrix \eqref{eq:sourcematrixtrans} vanishes~\cite{Kaminski:2009dh}.

\bibliographystyle{JHEP}
\bibliography{STSB-new}
\end{document}